\documentclass[pstricks,siunitx,addfonts,theorem,font=palatino]{tumphthesis}

\pdfoutput=1 
\newcommand{\nubar}{\ensuremath{\bar{\nu}}}
\newcommand{\nue}{\ensuremath{\nu_e}}
\newcommand{\numu}{\ensuremath{\nu_\mu}}
\newcommand{\nux}{\ensuremath{\nu_x}}
\newcommand{\nuebar}{\ensuremath{\bar{\nu}_e}}
\newcommand{\mean}[1]{\ensuremath{\langle #1 \rangle}}

\subject{Abschlussarbeit im Masterstudiengang Physik}
\title{Detecting Fast Time Variations in the Supernova Neutrino Flux with Hyper-Kamiokande}

\subtitle{\foreignlanguage{german}{Suche nach hochfrequenten Oszillationen des Supernovaneutrino-Flusses mit Hyper-Kamiokande}}
\author{Jost Migenda}
\renewcommand{\pdfauthor}{Jost Migenda}
\renewcommand{\pdftitle}{Detecting Fast Time Variations in the Supernova Neutrino Flux with Hyper-Kamiokande}
\date{2015-06-22}

\uppertitleback{{\em “However big you think supernovae are, they’re bigger than that.”}\\ --- Donald~Spector~\cite{Munroe2014}}
\lowertitleback{Themensteller: Prof.\ Dr.\ Lothar~Oberauer\\
Betreuer: Dr.\ habil.\ Georg~G.~Raffelt\\
\\
Datum des Kolloquiums: 8. Juli 2015
}

\begin{document}

\frontmatter
\maketitle
\chapter{Abstract}\label{ch-abstract}

For detection of neutrinos from galactic supernovae, the planned Hyper-Kamiokande detector will be the first detector that delivers both a high event rate (about one third of the IceCube rate) and event-by-event energy information. In this thesis, we use a three-dimensional computer simulation by the Garching group to find out whether this additional information can be used to improve the detection prospects of fast time variations in the neutrino flux.

We find that the amplitude of SASI oscillations of the neutrino number flux is energy-dependent. However, in this simulation, the larger amplitude in some energy bins is not sufficient to counteract the increased noise caused by the lower event rate.
Finally, we derive a condition describing when it is advantageous to consider an energy bin instead of the total signal and show that this condition is satisfied if the oscillation of the mean neutrino energy is increased slightly.
\setcounter{tocdepth}{2}
\tableofcontents

\mainmatter
\chapter{Introduction}\label{ch-introduction}

The history of supernova observation goes back nearly two millennia, with the first reports of a supernova written in the year 185 in ancient China. Later well-known reports are from 1054 (the supernova resulting in the Crab Nebula) and Kepler’s supernova in 1604. The assumption that these were new stars (from Latin \emph{nova}, meaning “new”) was finally rejected many centuries later. Today, it is overwhelmingly clear that supernovae originate from stars exploding at the end of their lifetime.

Baade and Zwicky in 1934 were the first to suggest that supernovae may be produced by stars at the end of their lifetime, leaving behind a neutron star~\cite{Baade1934}. Hoyle and Fowler in 1960 described two basic mechanisms of supernovae: the thermonuclear runaway ignition that is today associated with type Ia supernovae, as well as the implosion of a stellar core~\cite{Hoyle1960}. Early simulations by Colgate and others during the 1960s described the hydrodynamical behavior of the collapsing matter and the role of neutrinos in energy transfer~\cite{Colgate1961, Colgate1966}. The role of neutrinos in reheating the stalled shock front and thus causing the explosion was described by Wilson and Bethe in 1985~\cite{Wilson1982, Bethe1985}, thus completing the now generally accepted explosion mechanism of supernovae.

A core collapse supernova starts when the core of a massive star collapses because its self-gravitation has become stronger than the electron degeneracy pressure. As the density of the inner core grows rapidly, its equation of state stiffens. Infalling matter now hits a “wall” and is reflected, forming an outgoing shock wave, which uses up most of its energy to dissociate the nuclei and stagnates. Meanwhile, matter from outer layers of the star keeps falling in, creating a mostly stationary accretion shock front~(SAS).
In 2003, Blondin and others showed that small initial perturbations during this phase, like inhomogeneities in the infalling matter, can cause intense turbulence and sloshing motions of the shock front~\cite{Blondin2003}. This phenomenon has since been named “Standing Accretion Shock Instability” or SASI.\footnote{A simple yet very impressive demonstration of this phenomenon in shallow water --~humorously entitled SWASI~-- was described by Foglizzo and others~\cite{Foglizzo2012}.}

Due to a lack of observational data from the inner regions of supernovae, progress in supernova modeling has mostly come from computer simulations. With rapidly increasing computational power, simulations have just in recent years moved from predominantly two-dimensional, axially symmetric models to three-dimensional models with increasingly sophisticated treatment of energy-dependent neutrino transport and other subtle effects.
Since its discovery, SASI has been reproduced in many different two- and three-dimensional simulations and could play a potentially decisive role in triggering supernova explosions~\cite{Endeve2012, Hanke2013}. These simulations have shown that the sloshing motions of SASI cause time dependent oscillations of the rate of matter infall, which lead to oscillations of the neutrino number flux. They have also shown that SASI leads to time dependent oscillations of the mean neutrino energy.

The supernova 1987A in the Large Magellanic Cloud was the first and to date only supernova which was close enough to Earth for the associated neutrinos to be detected by several neutrino detectors around the world, making it one of only two identified astrophysical neutrino sources to date~-- the other one being, of course, the Sun.
The roughly two dozen events --~twelve in the Kamiokande detector~\cite{Hirata1987, Hirata1988}, eight in the IMB detector~\cite{Bionta1987} and five in the Baksan detector~\cite{Alekseev1987}~-- displayed an average energy and time distribution which are in agreement with the generally accepted explosion mechanism. However, the data did not enable us to deduce details of the mechanism, many of which remain contested until today.

The observational landscape today has improved dramatically since 1987:
The IceCube detector at the South Pole uses a \SI{1}{km^3} large volume of ice as detector material. Neutrinos interacting with the hydrogen nuclei in the ice produce positrons through inverse beta decay on hydrogen nuclei:
\begin{equation*}
p + \nuebar \rightarrow n + e^+
\end{equation*}
The typical energy of supernova neutrinos is $\mathcal{O}(\SI{10}{MeV})$ and thus much higher than the positron mass. The positrons are therefore highly relativistic and produce Cherenkov radiation, which is then detected by optical modules embedded in the ice. Due to its large volume, IceCube could deliver a very high event rate of $\mathcal{O}(\SI{e6}{Hz})$ for a galactic supernova. On the other hand, the large distance between detector modules means that IceCube will pick up at most one photon from any given neutrino interaction. It is unable to differentiate these events from background and will therefore need to rely on an increase of the event rate across all of its modules to identify a supernova.

The Super-Kamiokande detector in Japan is a cylindrical tank in a mine about \SI{1000}{m} underground. It relies on inverse beta decay, similar to IceCube, but has a much higher density of detector modules, which makes it possible to reconstruct individual events and measure the positron energy. Super-Kamiokande would be able to see an essentially background-free signal, albeit with a much lower event rate of $\mathcal{O}(\SI{e4}{Hz})$.

\enlargethispage{-\baselineskip} 
The planned Hyper-Kamiokande detector is about 25 times larger than its predecessor and will, once it is built, combine the best of both worlds, offering a high event rate of about one third the IceCube rate, while providing the same very low background and event-by-event energy information as its predecessor.
Depending on its distance from Earth, the next supernova in the Milky Way could cause between $10^6$ and $10^8$ events in these and a number of other detectors.


Lund and others were the first to analyze the power spectrum of SASI oscillations and their detectability~\cite{Lund2010}.
More recently, Tamborra and others have shown that, using the IceCube detector, it is possible to see the oscillations of the neutrino number flux caused by SASI for a range of supernova scenarios~\cite{Tamborra2013, Tamborra2014}.

In this thesis, we will look at the planned Hyper-Kamiokande detector and will investigate whether its event-by-event energy information can help in discovering SASI oscillations in the neutrino signal from future galactic supernovae. This could lead to an experimental confirmation of the picture of the supernova explosion mechanism which is currently mostly based on computer simulations. It could also help clarify details of the mechanism, where current simulations give conflicting results.

We will start by discussing the established knowledge of supernova and neu\-trino physics in chapters~\ref{ch-supernovae} and~\ref{ch-neutrinos}, respectively.
In chapter~\ref{ch-data}, we will describe our data set and analysis techniques.
In chapter~\ref{ch-or}, we will discuss the relative detection prospects of the IceCube and Hyper-Kamiokande detectors depending on the supernova distance. We will also investigate whether the amplitude of the SASI oscillations is a function of neutrino energy and whether the SASI detection probability can thus be improved by focussing on an appropriate energy range.
Finally, we will conclude in chapter~\ref{ch-conclusionsoutlook} and give an outlook on future research directions.

\chapter{Core-Collapse Supernovae}\label{ch-supernovae}

\section{Classification of Supernovae}

\subsection{Spectral Classification}

A classification of supernovae into two types, based on their spectral lines, was introduced by Minkowski in 1941~\cite{Minkowski1941}. Zwicky later refined this classification into five types, taking into account also the shape of the light curve~\cite{Zwicky1964}.

\begin{figure}[htbp]
	\centering
	\includegraphics[scale=1.2]{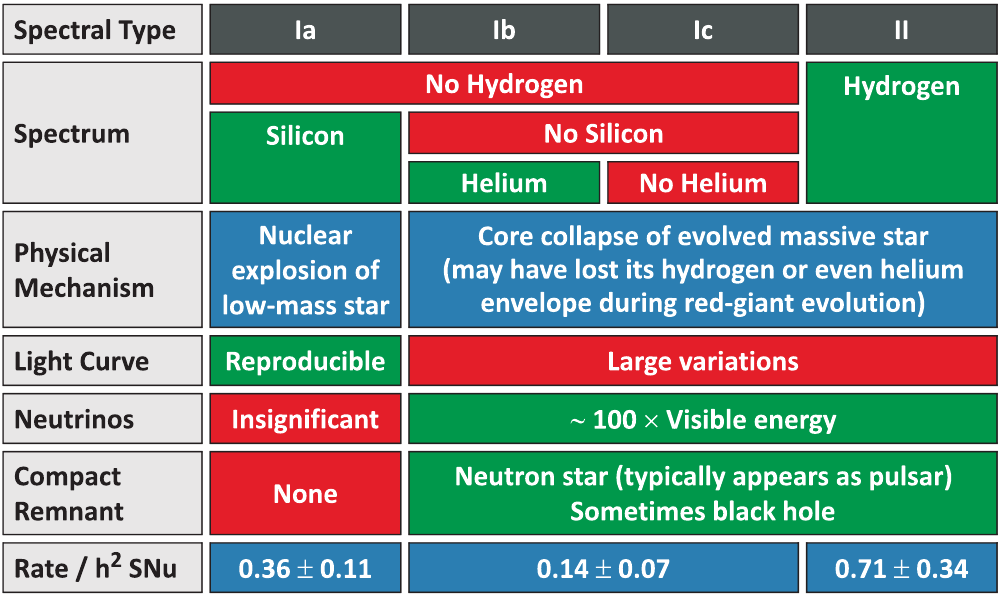}
	\caption{Overview of spectral types and some of their properties. In total, more than 6800 supernovae have been detected until June 2015~\cite{Asiago2015}. (Figure taken from reference~\cite{Raffelt2011}.)}
	\label{fig-sn-classification}
\end{figure}

Today, supernovae are divided into two basic types: type~I, if their spectrum contains no hydrogen lines, or type~II, if it does. Both are divided into subtypes, depending on the presence or absence of additional lines in their spectra. An overview is given in figure~\ref{fig-sn-classification}.

These spectral types are sometimes divided into additional subtypes like type~IIP (supernovae with a plateauing light curve) or type~IIL (supernovae with a light curve that falls off linearly). In addition, some supernovae can be sorted into one of these categories but feature some peculiar properties~\cite{Gonzalez-Gaitan2014,Maeda2009,Ouyed2013,Kleiser2011}. While suggestions abound, the physical reasons are often unclear.

It is generally accepted, however, that these spectral properties are mostly de\-ter\-mined by the outer layers of the progenitor~-- for example by the size of its hydrogen envelope and by its $^{56}$Ni content~\cite{Heger2003}. Therefore, different spectral properties do not generally correspond to different explosion mechanisms.

\subsection{Classification by Explosion Mechanism}
It is also possible to classify supernovae based not on their spectra, but on the properties of their progenitor and thus on their explosion mechanism.

\subsubsection{Thermonuclear Supernovae}
It was first suggested in 1960 by Hoyle and Fowler that supernovae of type~Ia in the above classification system are caused by thermonuclear explosions~\cite{Hoyle1960}. Whelan and Iben later proposed a progenitor model~\cite{Whelan1973}, which has since been refined into the model that is widely used today. In this model, thermonuclear supernovae originate in gravitationally bound systems of two stars, one of which is a white dwarf consisting primarily of carbon and oxygen. If both stars are close enough, the white dwarf will over time accrete matter from its companion, until its mass reaches the Chandrasekhar mass limit. During accretion, the nuclear fuel in its core heats up until a runaway fusion process starts, which releases $\mathcal{O}(\SI{e51}{erg})$ within seconds.

A large part of this energy is released in the form of kinetic energy, ripping the white dwarf apart and expelling its companion star. Neutrinos, which are created mostly through electron capture on free protons and other nuclei, i.\,e.
\begin{equation*}
e^- + (A,Z) \rightarrow (A, Z-1) + \nu_e,
\end{equation*}
play only a minor role, being responsible for less than 10\,\% of the total energy release~\cite{Iwamoto2006}.

While this mechanism is widely accepted, it is not without problems and many authors have used observations of individual supernovae (such as SN2006gz~\cite{Hicken2007}) as well as observations of star populations~\cite{Pritchet2008} to argue that differing mechanisms, such as the collision of two white dwarfs, may be at work for some type Ia supernovae.\footnote{For an overview of open questions and as yet unexplained observations, see reference~\cite{Janka2011}.}

\subsubsection{Core-Collapse Supernovae}
In contrast to thermonuclear supernovae, whose progenitors are rather light white dwarfs, core-collapse supernovae have heavy stars with more than about eight solar masses as progenitors. In the core of such a star, temperatures and densities are high enough for nuclear fusion to produce heavy elements up to iron. In the late stages, the star has a core consisting mostly of iron and nickel. Once the mass of this core crosses the Chandrasekhar limit, it starts to collapse and the core’s density increases until it reaches nuclear density. At this point, its equation of state stiffens. The infalling matter now bounces on the core and is reflected outwards as a shock wave. When this shock wave is reheated by neutrinos, the outer shells of the star are expelled and the inner core leaves either a neutron star or a black hole as a remnant.

While this basic mechanism is believed to be the same for most core-collapse supernovae, their outer appearance can vary widely. Depending on the progenitor’s properties, core-collapse supernovae therefore correspond to any of the spectral types Ib, Ic and II.

Approximately 99\,\% of the energy, roughly \SI{3e53}{erg}, is released in the form of neutrinos. These are produced in the inner regions of the star and will traverse the outer layers nearly unhindered. Measuring the neutrino flux and average energy therefore allows us to study the processes at work during the explosion.
In the remainder of this thesis, we will focus on core-collapse supernovae.

\section{Phases of a Core-Collapse Supernova}\label{ch-sn-phases}

The progenitor of a core-collapse supernova is a massive star, that has completed all phases of nuclear fusion in its core, which now consists of iron and nickel. Nuclear burning is still ongoing in the outer shells, where additional iron is produced and sinks down to the core, adding to its mass. Following figure~\ref{fig-sn-phases-overview}, the “delayed explosion mechanism” of a core-collapse supernova will progress in six phases:

\begin{figure}[p]
	\centering
	\includegraphics[scale=0.94]{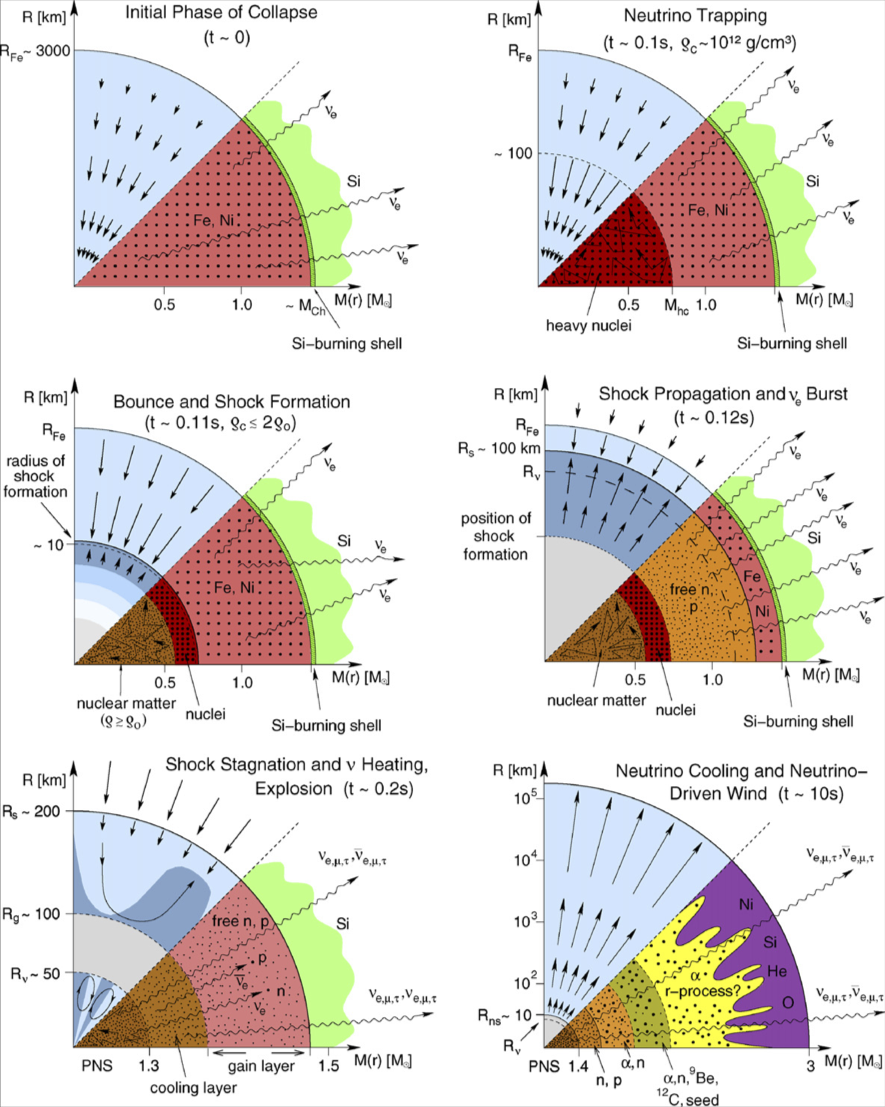}
	\caption{Sketch of the six phases of the delayed explosion mechanism as described in the text. In each panel, the upper section shows the dynamical processes, with arrows representing velocity vectors, while the lower section shows the nuclear composition of the star. (Figure taken from reference~\cite{Janka2007}.)}
	\label{fig-sn-phases-overview}
\end{figure}

\begin{enumerate}
\item \emph{Initial phase:} Once the mass of the core reaches about $M_\text{Ch} \approx 1.4\ M_\odot$, the self-gravitation is stronger than the electron degeneracy pressure. The core becomes unstable and starts to collapse, leading to electron capture in nuclei, with the resulting neutrinos transporting energy away from the core. This process reduces the pressure and therefore accelerates the collapse.

\item \emph{$\nu$ trapping:} After about \SI{100}{ms}, the inner core reaches a density of about $\SI{e12}{g/cm^3}$. At this density, neutrinos are trapped by coherently enhanced elastic scattering, since their mean free path is smaller than the radius of the inner core. It will take them a fraction of a second to escape.

\item \emph{Bounce and shock formation:} After about \SI{110}{ms}, the density of the inner core surpasses nuclear density, reaching about $\SI{3e14}{g/cm^3}$. At this point, the equation of state of the inner core stiffens. Infalling matter now hits a “wall” and is reflected, resulting in an outgoing shock wave.

\item \emph{Shock propagation:} After about \SI{120}{ms}, the shock wave reaches the outside of the iron core, dissociating the iron nuclei into free nucleons. Since the electron capture rate on free protons is much higher than on the larger, neutron-rich nuclei, this leads to a sudden increase in the reaction
\begin{equation*}
e^- + p \rightarrow n + \nu_e.
\end{equation*}
Since the matter density in the outer parts of the core is not high enough to trap the neutrinos, a sudden \nue\ burst is released.

\item
\emph{Shock stagnation and $\nu$ heating:} After about \SI{200}{ms}, the shock wave stagnates at a radius of about \SIrange{100}{200}{km}, having used up most of its energy to dissociate heavy nuclei into their constituent nucleons. The shock front now becomes mostly stationary, while infalling matter from outer layers creates an accretion shock, powering neutrino emission. At this phase, convection sets in at the accretion shock layer.

Meanwhile, neutrinos are starting to escape from the inner core, depositing energy in the material behind the shock front mainly by neutrino capture on free nucleons:
\begin{eqnarray*}
\nubar_e + p \rightarrow n + e^+\\
\nu_e + n \rightarrow p + e^-
\end{eqnarray*}
This heating increases the pressure in the region behind the shock front and reignites the shock wave, leading to the explosion that expels the matter in the outer shells of the star.

\item \emph{$\nu$ cooling:} During the final phase, which has a duration of about ten seconds, the remnant of the core, a proto-neutron star (PNS), cools by diffusive neutrino transport.
\end{enumerate}

\begin{figure}[ht]
	\centering
	\includegraphics[scale=0.85]{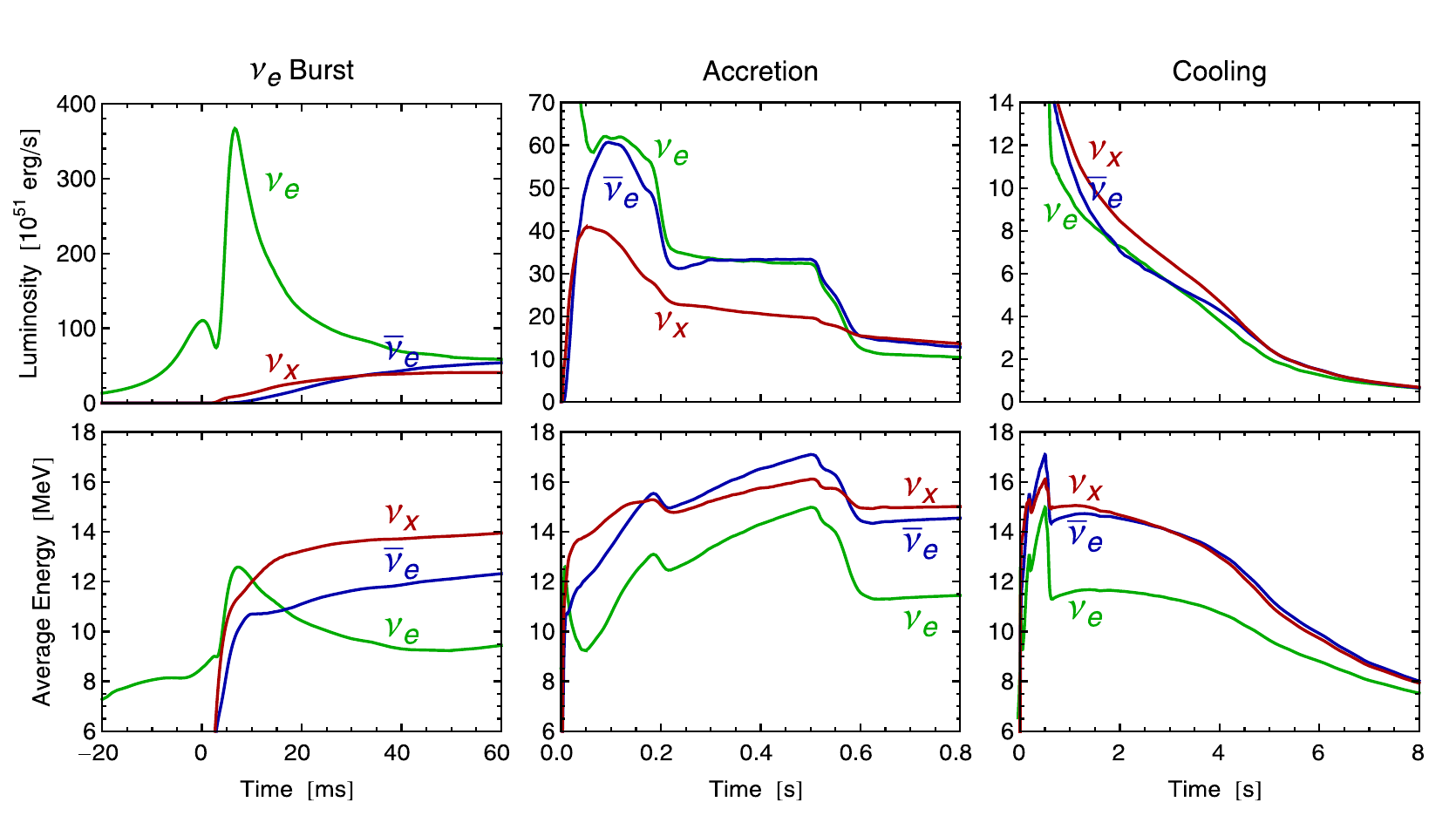}
	\caption{Luminosity (upper panels) and mean energy (lower panels) of \nue\ (green), \nuebar\ (blue) and \nux\ (red) for a spherically symmetric model by the Garching group. From left to right, the panels show the prompt \nue\ burst, the following phase of shock stagnation, and finally the neutrino cooling. (Figure taken from reference~\cite{JUNO2015}.)}
	\label{fig-sn-lum-meane}
\end{figure}

Throughout this process, $\nu$ emission happens in three distinct steps, which can be easily identified. (See figure~\ref{fig-sn-lum-meane}.)

\begin{figure}[htbp]
	\centering
	\includegraphics[scale=0.5]{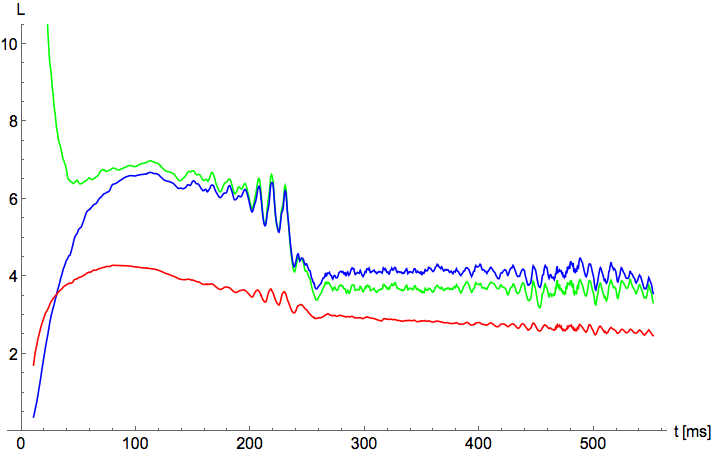}
	\caption{Luminosity (in arbitrary units) of \nue\ (green), \nuebar\ (blue) and \nux\ (red) for the data set described in chapter~\ref{ch-data-dataset}. It is easy to identify the \nue\ burst and the following phase of shock stagnation including the SASI oscillations at about \SI{200}{ms}.}
	\label{fig-sn-luminosity}
\end{figure}

\begin{figure}[hbtp]
	\centering
	\includegraphics[scale=0.5]{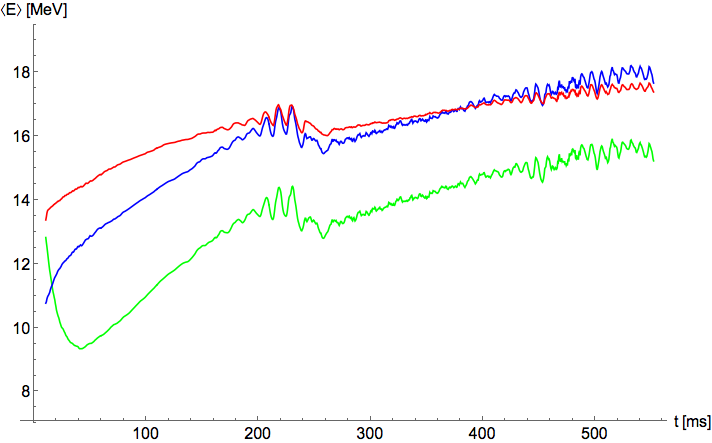}
	\caption{Mean energy (in \si{MeV}) of \nue\ (green), \nuebar\ (blue) and \nux\ (red) for the data set described in chapter~\ref{ch-data-dataset}. At about \SI{200}{ms}, we can identify the SASI oscillations.}
	\label{fig-sn-mean-energy}
\end{figure}

The first step is a prompt \nue\ burst from electron capture, coinciding with phase 4 above. With a duration of roughly \SI{10}{ms}, this is expected to give a very sharp and unmistakeable feature of nearly pure \nue, which is easily identifiable in figures~\ref{fig-sn-luminosity} and~\ref{fig-sn-mean-energy}. Neither \nuebar, nor neutrinos or antineutrinos of other flavors (referred to as \nux) are expected in this phase. Since this signal originates in the iron core, which collapses at a well-defined set of physical conditions, independent of the properties of the outer shells of the star, it has turned out to be nearly unchanged for a variety of different progenitors~\cite{Kachelries2005}.

The second step has a duration of several \SI{100}{ms} and coincides with the shock stagnation in phase 5 above. In this phase, the luminosities $L_{\nu_e}$ and $L_{\nubar_e}$ are roughly equal (and higher than $L_{\nu_x}$) while the average energies are unequal ($\langle E_{\nu_e} \rangle < \langle E_{\nubar_e} \rangle \approx~\langle E_{\nu_x} \rangle$), which contributes to the de-leptonization of the core. During this phase, the oscillations caused by SASI can be seen very clearly in figure~\ref{fig-sn-luminosity} and somewhat less clearly in figure~\ref{fig-sn-mean-energy}.

The last step is the $\nu$ cooling (phase 6), which has a duration of roughly ten seconds. In this phase, the supernova remnant cools through diffusive $\nu$ emission, whose composition is governed by a number of different physical processes, including nucleon-nucleon bremsstrahlung and neutrino-antineutrino pair annihilation~\cite{Hannestad1998,Thompson2000,Buras2003,Keil2003}. In this phase, the luminosities of the neutrino species are roughly equal, $L_{\nu_e} \approx L_{\nubar_e} \approx L_{\nu_x}$, all falling off exponentially with time, while the average energies remain unequal.


\section{Computer Simulations of Core-Collapse Supernovae}

Attempts to develop an understanding of the explosion mechanism of core-collapse supernovae have been hindered by several factors:

\begin{itemize}
\item As is frequently the case in astrophysics, experiments that try to reproduce the necessary conditions in a laboratory are simply not feasible.
\item Neutrino astrophysics of supernovae is an extremely data-starved field, with a total of about two dozen events from SN1987A.
\item Data from optical observations of supernovae is far more widely available, but of limited use: While it can inform us about many aspects of the outer layers of the star, like their chemical composition, it gives very little insight into the physical processes in the core at the moment of explosion. This is due to the fact that, while the neutrino signal originates at the core at the instant of the explosion, the photon signal originates when the shock wave of the explosion hits the star’s photosphere.\footnote{Accordingly, the optical signal originates a few hours or days --~depending on the size and composition of the star~-- after the neutrino signal. Since the neutrinos are highly relativistic, this time difference is preserved on the way to Earth, which has motivated the creation of the SuperNova Early Warning System (SNEWS)~\cite{Scholberg2000, Antonioli2004}. SNEWS will notify subscribers in near-real time about coincident neutrino bursts in the represented neutrino detectors, which are likely to signify a galactic supernova. Since some of the detectors are able to give directional information, this would allow astronomers to position their telescopes accordingly, thus potentially enabling them to measure the supernova light curve at a very early stage, which is hardly possible by other means. Additionally, SNEWS could inform other experiments that might not be able to trigger on the supernova by themselves, to ensure that the data of those experiments gets scrutinized heavily and stored permanently~\cite{Scholberg2000}.}
\end{itemize}

In the last decades, computer simulations of supernovae are an area which has made remarkable progress, despite these factors. This was caused by a dramatic increase in available computing resources as well as by notable increases in our knowledge regarding many of the physical processes involved in the explosion of a core-collapse supernova --~whether it be various aspects of neutrino physics or nuclear fusion cross sections.

Pioneering contributions to numerical modeling of supernova explosions were made by Colgate\,\&\,White~\cite{Colgate1966}, Arnett~\cite{Arnett1967} and Wilson~\cite{Wilson1971,Wilson1982}. Despite some exceptions, these early simulations were usually one-dimensional, imposing spherical symmetry on the progenitor and the resulting processes~\cite{Janka2012}. Optical observations of SN1987A made it obvious that this assumption is not fulfilled in nature~\cite{Hillebrandt1989}, increasing the interest in more complex two-dimensional simulations. Today, two-dimensional simulations are still an area of ongoing research. Only in very recent years has it become feasible to simulate three-dimensional models including detailed physical processes for neutrino creation and transport~\cite{Janka2012, Hanke2013}.

The difficulty in simulating supernova explosions stems both from the huge computation power required, as well as complex physical problems: Supernova modelling stands out from most other fields of research in that it involves all known fundamental forces~-- gravity as well as the strong and electroweak force. It also includes non-linear hydrodynamics, relativistic effects and extreme conditions, which are sometimes not well known from laboratory experiments. The relativistic effects are currently often treated as a modified potential in Newtonian gravity to simplify calculations, which might cause an error of some tens of percent in some physical quantities and lead to a markedly different outcome~\cite{Muller2012}.

In addition, it is not always obvious whether some property of a simulation is an actual physical effect or just an artifact of the limitations inherent in the computer model.

For example, the role of SASI oscillations in supernovae recently was the subject of intense debate: While these oscillations were seen in a number of two-dimensional simulations, early three-dimensional simulations did not show this phenomenon, leading some authors to conclude that it had to be an artifact of the restriction of rotational symmetry imposed on earlier simulations~\cite{Burrows2012}. More recently, other authors have found SASI oscillations in their three-dimensional simulations (including the one used in this thesis, see figures~\ref{fig-sn-luminosity} and~\ref{fig-sn-mean-energy}), continuing the debate~\cite{Hanke2013}.

As another example, progenitors in computer simulations often do not explode on their own but need to be ignited artificially. In fact, the first successful neutrino-driven explosion in a three-dimensional simulation was only achieved very recently by the Garching group~\cite{Melson2015}. It is still unclear whether the lack of explosions shows flaws in our modeling of the known physical processes, or whether it points to new physics --~e.\,g. new particles, like axions or dark sector particles~\cite{Turner1988, Janka1996,Dreiner2014,Kazanas2015}~--, which are completely missing from current models.

\chapter{Neutrino Physics and Detection}\label{ch-neutrinos}

Among the elementary particles that are known to exist, neutrinos are, without a doubt, the most mysterious ones. Even today, 85 years after their original postulation and nearly 60 years after the first experimental confirmation, they still pose many questions and are considered to be among the most promising pointers to physics beyond today’s standard model of particle physics.

In this chapter, we will start by discussing the history of neutrino physics and their role in modern particle physics, with special focus on neutrino oscillations.
We will then go on to discuss general principles of neutrino detection through inverse beta decay. Finally, we will describe in detail the current IceCube detector and the planned Hyper-Kamiokande detector as well as their respective predecessors.

\section{History of Neutrino Physics}

It was already known in the 1920s that some nuclei undergo beta decay, expelling an electron:
\begin{equation*}
^A_Z X \rightarrow\ ^{\quad A}_{Z+1}X' + e^-\ \ (+ \nuebar )
\end{equation*}
Energy measurements of the end products revealed that the electron energy spectrum was continuous, as opposed to a discrete line, which one would expect in a two-body decay if energy is conserved. While Bohr considered the possibility that conservation of energy might not be valid in this situation~\cite{Bohr1932}, Pauli in 1930 proposed an alternative solution to this puzzle, by introducing a new particle~-- the neutrino\footnote{Pauli originally called the proposed new particle “neutron”. After the discovery of what is today called the neutron by Chadwick in 1932~\cite{Chadwick1932}, Pauli’s proposed particle was renamed “neutrino”. This contraction of the Italian word “neutronino” --~meaning “little neutron”~-- was jokingly suggested by Amaldi~\cite{Amaldi1998,Bonolis2005} and was later popularized by Fermi~\cite{Fermi1934d,Fermi1934i}.}~\cite{Pauli1985}.

\enlargethispage{\baselineskip}
In the following years, work on the theoretical description of this new particle progressed relatively quickly. In 1934, Fermi published a theoretical description of beta decay~\cite{Fermi1934d}, which included the interaction Hamiltonian of beta decay and a discussion of the dependence of the beta decay end point spectrum on the neutrino mass.

Experimental successes on the other hand were still far off. An early estimation of the neutrino-nucleon scattering cross section by Bethe and Peierls resulted in $\sigma < \SI{e-44}{cm^2}$ for a \SI{2.3}{MeV} neutrino beam, leading them to the conclusion “that there is no practically possible way of observing the neutrino”, even for higher energies~\cite{Bethe1934}.
Several experiments during the next two decades delivered in\-crea\-sing\-ly strong evidence for missing momentum in beta decay~\cite{Leipunski1936,Crane1938,Jacobsen1945,Christy1947,Rodeback1952}, but non-conservation of energy was at the time still considered a possibility~\cite{Fukugita2003}.

Finally, Reines and Cowan used a liquid scintillator detector in an attempt to detect neutrinos through inverse beta decay ($\nuebar + p \rightarrow n + e^+$), without the need to assume conservation of energy and momentum~\cite{Reines1953,Cowan1953}. In 1953, using the Hanford nuclear reactor as a neutrino source, they found a tentative signal, which was not yet conclusive~\cite{Reines1953a}. Three years later, using the Savannah River nuclear reactor, they were able to confirm this excess and thus demonstrate the existence of neutrinos~\cite{Reines1956,Cowan1956}.

A few years before, several experiments measuring the lifetime of double beta decay had already suggested that neutrinos and antineutrinos are distinct particles with different physical properties~\cite{Konopinski1953}. More evidence was found in the mid-1950s by Davis~\cite{Davis1955}, who used an antineutrino source to look for evidence for the reaction
\begin{equation*}
^{37}\text{Cl} + \nuebar \rightarrow\ ^{37}\!\text{Ar} + e^-\!.
\end{equation*}
Within three years, Davis reached a sensitivity of $0.05$ times the theoretical cross section of the equivalent neutrino reaction
\begin{equation*}
^{37}\text{Cl} + \nue \rightarrow\ ^{37}\!\text{Ar} + e^-
\end{equation*}
without finding evidence for the reaction, thus conclusively demonstrating that neutrinos and antineutrinos behave differently~\cite{Davis2003}.

Around the same time, the neutrino’s parity and helicity were measured in intri\-cate experiments by Wu~\cite{Wu1957} and Goldhaber~\cite{Goldhaber1958} and their respective collaborators.

In 1962, it was similarly concluded that muon neutrinos are distinct from electron neutrinos. This was done by producing $\nu_\mu$ in pion decay through $\pi^\pm \rightarrow \mu^\pm + (\nu_\mu / \nubar_\mu)$ and observing that these neutrinos produced muons, rather than electrons, in the detector~\cite{Danby1962}.

After the discovery of the $\tau$ lepton in 1975~\cite{Perl1975}, the existence of a corresponding neutrino species was generally expected. Experimental evidence for the existence of $\nu_\tau$ was finally found in 2000 by the DONUT experiment at Fermilab~\cite{Kodama2001}.
Experimental measurements of the $Z$ boson’s decay width at LEP show that there are no additional light neutrinos~\cite{PDG2014}.

\section{Neutrino Oscillations}

\subsection{The Solar Neutrino Problem}
For centuries, the source of the energy generated in the Sun had been a mystery. Early suggestions like chemical or gravitational energy would have given the Sun a life time of several thousand or several million years, which appeared irreconcilable with the discovery of fossils or rocks that appear to be several billions of years old.\footnote{An account of the discussion between Kelvin and contemporary geologists is given in reference~\cite{England2007}.}

\begin{figure}[ht]
	\centering
	\includegraphics[scale=0.8]{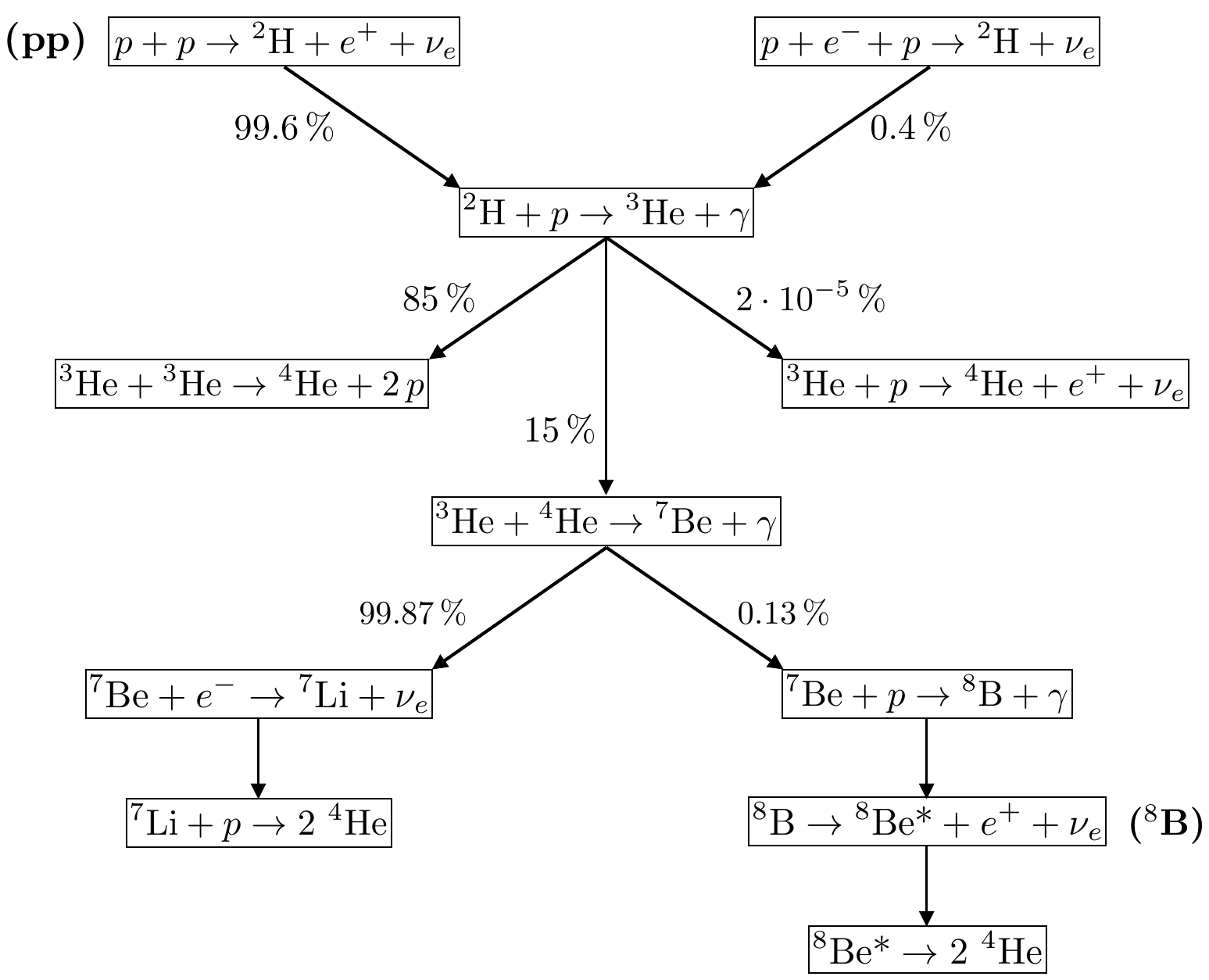}
	\caption{Reactions and branching ratios for the pp chain~\cite{Bilenky1999}.}
	\label{fig-nu-pp}
\end{figure}

\begin{figure}[htb]
	\centering
	\includegraphics[scale=0.85]{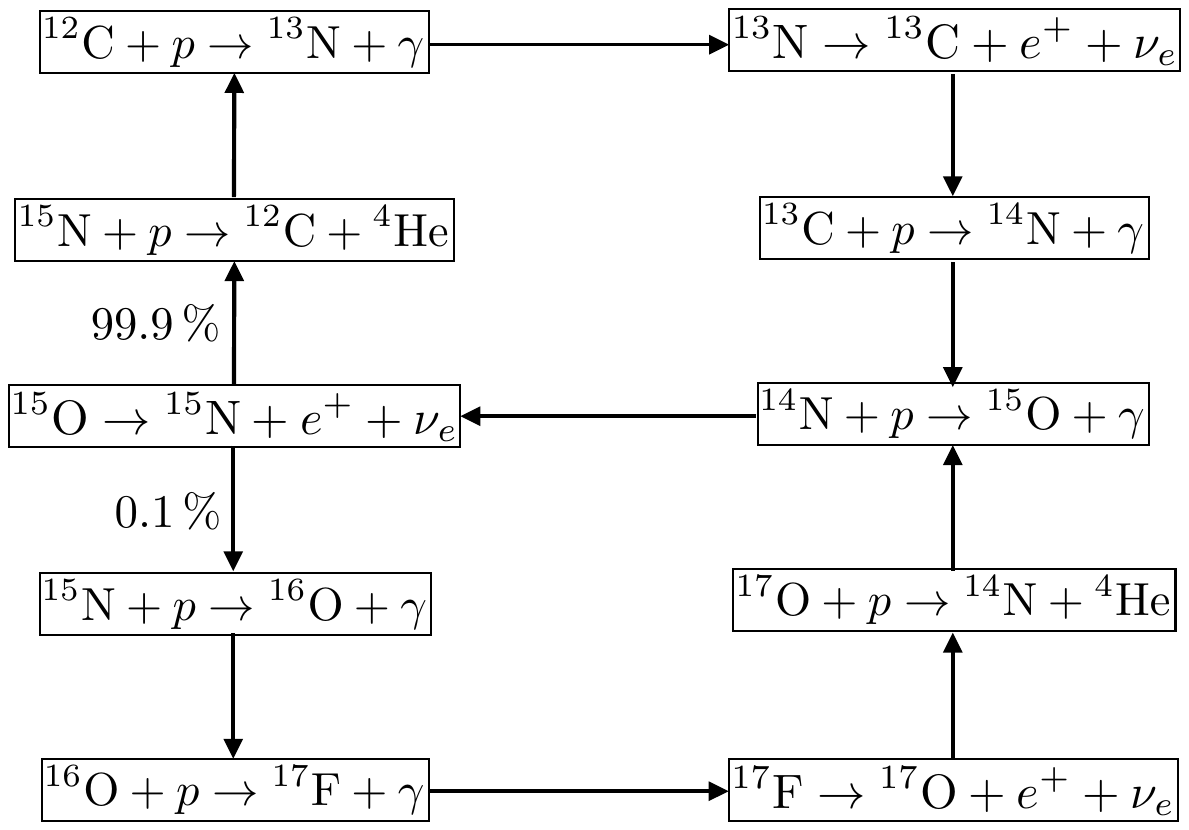}
	\caption{Reactions and branching ratios for the CNO cycle~\cite{Bilenky1999}.}
	\label{fig-nu-cno}
\end{figure}

Nuclear fusion as a stellar energy source was suggested by Eddington in 1920~\cite{Eddington1920}. In 1939, Bethe expanded upon Eddington’s proposal and described two reaction chains that are responsible for energy generation in the Sun: the pp chain (which fuses four protons into a $^4$He nucleus) and the CNO cycle (which does the same using a carbon nucleus as a catalyst)~\cite{Bethe1939}. In both processes, the mass difference between the initial and the final material is converted into neutrinos or photons with energies of $\mathcal{O}$(\si{MeV}). Both reaction chains are displayed in figures~\ref{fig-nu-pp} and~\ref{fig-nu-cno}.

The interior of the Sun (like the interior of a supernova) is inaccessible to optical observations because the mean free path of a photon is much smaller than the radius of the Sun. It is therefore impossible to observe the photons generated in these reactions in the Sun’s interior. Neutrinos with $\mathcal{O}$(\si{MeV}) energies, on the other hand, can escape nearly unhindered. If we are able to detect these solar neutrinos, we could observe the nuclear fusion processes inside the Sun directly.

The importance of the reaction
\begin{equation*}
^3\text{He} +\/ ^4\text{He} \rightarrow\ ^7\text{Be} + \gamma
\end{equation*}
was underestimated at first. Laboratory experiments in 1959 found that its cross section was two orders of magnitude larger than previously thought~\cite{Holmgren1959}. Accordingly, the reaction chain
$$
^7\text{Be} + p \rightarrow\ ^8\text{B} + \gamma \rightarrow\ ^8\text{Be*} + e^+ + \nue
$$
produces a neutrino flux that is much higher than previously expected. These $^8$B neutrinos can have an energy of more than \SI{10}{MeV}, making them easier to detect than neutrinos from other reactions which --~although some of them are more abundant~-- have a lower energy that is below the energy threshold required by easily available detector materials.

Following this discovery, Bahcall and Davis in 1964 proposed an experiment using \SI{380000}{l} of perchlorethylene, C$_2$Cl$_4$, to look for the reaction
$$
^{37}\text{Cl} + \nue \rightarrow\ ^{37}\!\text{Ar} + e^-\!,
$$
which can be caused by $^8$B neutrinos~\cite{Bahcall1964,Davis1964}. This reaction is advantageous due to its low threshold energy of about \SI{0.8}{MeV}, the easily available detector material and an enhanced cross section due to three excited states of $^{37}$\!Ar~\cite{Bahcall1964}.
Such an experiment was expected to provide first direct evidence that nuclear fusion is indeed the source of the energy generated in the Sun. In addition, since the reaction
$$
^7\text{Be} + p \rightarrow\ ^8\text{B} + \gamma
$$
has a strong temperature dependence, a measurement of the $^8$B neutrino flux can be used to measure the temperature inside the Sun.

The detector was built in the Homestake Gold Mine in South Dakota, \SI{1480}{m} underground --~corresponding to \num{4400} meters of water equivalent (m.w.e.)~-- to reduce the background from cosmic ray muons.

First results published in 1968 found no evidence of solar neutrinos. By comparing the resulting upper limit on the solar neutrino flux to theoretical predictions, it was deducted that the CNO cycle is responsible for less than 9\,\% of the total energy generation in the Sun. The best theoretical prediction for the expected neutrino flux, however, was about twice as high as the experimental upper limit~\cite{Bahcall1968,Davis1968}.

In the mid-1980s, the Kamiokande experiment (see chapter~\ref{ch-nu-sk}) began measuring solar neutrinos and similarly found an event rate that is about half the theoretically predicted rate~\cite{Fukuda1996}.
In the 1990s, a number of experiments began using gallium as a detector material to look for solar neutrinos using the reaction
$$
^{71}\text{Ga} + \nue \rightarrow\ ^{71}\text{Ge} + e^-\!,
$$
which has an energy threshold of about \SI{0.2}{MeV}. This enables the detection of pp neutrinos, which are by far the most abundant type of solar neutrinos. As we see from figure~\ref{fig-nu-pp}, their flux is nearly four orders of magnitude higher than that of $^8$B neutrinos.

The experiment GALLEX (GALLium EXperiment) and its successor GNO (Gallium Neutrino Observatory) at the underground laboratory Gran Sasso, as well as SAGE (Soviet-American Gallium Experiment) in the Baksan laboratory in the Caucasus region, all measured the solar neutrino flux and found values that are a little over half the theoretically predicted rate~\cite{Hampel1999,Altmann2000,Abdurashitov1999}.

This decade-long discrepancy between theoretical predictions and experimental measurements of the solar neutrino flux became known as the “solar neutrino problem”.
Experimental error as a source for this discrepancy seemed unlikely, given that a number of very different experiments had found compatible results independently. Another possible solution, mistakes in the standard solar model, seemed improbable as well, given the excellent agreement of that model with helioseismological measurements. It therefore appeared likely that a solution to this problem was to be found in particle physics. Among the number of suggested solutions, neutrino oscillations were a favorite of many physicists.

\subsection{Neutrino Oscillations}
To explain experiments that appeared to violate the weak charge universality, in 1963, Cabibbo proposed a new description of leptonic decays of strongly interacting particles~\cite{Cabibbo1963}.
Re-expressed in terms of the quark model, we write $u$ and $c$ for the eigenstates of the up and charm quarks in both the strong interaction and the weak interaction. For the down and strange quarks, we then find that the weak interaction eigenstates $d'$ and $s'$ are not equal to the strong interaction eigenstates $d$ and $s$. The relation between both sets of eigenstates is given by the linear combination
\begin{equation}\label{eq-nu-cabibbo}
\begin{pmatrix}
	d'\\
	s'
\end{pmatrix}
=
\begin{pmatrix}
	\cos{\theta_C} & \sin{\theta_C}\\
	- \sin{\theta_C} & \cos{\theta_C}
\end{pmatrix}
\begin{pmatrix}
	d\\
	s
\end{pmatrix}.
\end{equation}
The mixing angle $\theta_C \approx 0.2$ is called the Cabibbo angle.

In 1973, Kobayashi and Maskawa extended this description to include the third generation of quarks~\cite{Kobayashi1973}. The resulting $3 \times 3$ matrix is called the Cabibbo-Kobayashi-Maskawa matrix or CKM matrix.

Similarly, in the neutrino sector, the weak interaction eigenstates or flavor eigenstates \nue, $\nu_\mu$ and $\nu_\tau$ are distinct from the mass eigenstates $\nu_1$, $\nu_2$ and $\nu_3$. The flavor eigenstates can be written as a linear combination of the mass eigenstates using the Pontecorvo-Maki-Nakagawa-Sakata matrix or PMNS matrix,
\begin{equation}
\begin{pmatrix}
	\nue\\
	\nu_\mu\\
	\nu_\tau
\end{pmatrix}
=
\begin{pmatrix}
	c_{12} c_{13} & s_{12} c_{13} & s_{13} e^{- i \delta}\\
	-s_{12} c_{23} - c_{12} s_{13} s_{23} e^{i \delta} & c_{12} c_{23} - s_{12} s_{13} s_{23} e^{i \delta} & c_{13} s_{23}\\
	s_{12} s_{23} - c_{12} s_{13} c_{23} e^{i \delta} & -c_{12} s_{23} - s_{12} s_{13} c_{23} e^{i \delta} & c_{13} c_{23}
\end{pmatrix}
\begin{pmatrix}
	\nu_1\\
	\nu_2\\
	\nu_3
\end{pmatrix},
\end{equation}
where $s_{ij} = \sin{\theta_{ij}}$, $c_{ij} = \cos{\theta_{ij}}$ and $\delta$ is a CP-violating phase.

Pontecorvo in 1957 first discussed the possibility of neutrino oscillations~\cite{Pontecorvo1958}.
In 1962, Maki, Nakagawa and Sakata introduced a two-dimensional mixing matrix to describe the relationship between “weak neutrinos” (\nue\ and $\nu_\mu$) and “true neutrinos” (which are given by linear combinations of the former, similar to equation~\eqref{eq-nu-cabibbo})~\cite{Maki1962}.
A relation between neutrino oscillations in a two-flavor scenario and the solar neutrino flux measurements was then hypothesized by Pontecorvo in 1967~\cite{Pontecorvo1967}.

Neutrinos are produced in the Sun in the flavor eigenstate \nue. On the way between Sun and Earth, the three constituent mass eigenstates can then propagate indepen\-dent\-ly from each other.\footnote{This requires, of course, that the mass eigenstates are independent. In particular, at least two of the three eigenstates need to have a finite mass.} When transformed back into flavor eigenstates in a detector on Earth, a solar neutrino could interact as any one of the three weak interaction eigenstates.

\enlargethispage{-\baselineskip} 
While other experiments had already found evidence of neutrino oscillations~\cite{Fukuda1998}, the direct experimental confirmation came in 2002 from the SNO (Sudbury Neutrino Observatory) experiment. SNO is located in Canada’s Ontario province in a mine shaft at about \SI{2000}{m} (about \num{6000}\,m.w.e.) below the Earth’s surface. The location is among the deepest underground laboratories in the world and provides an excellent shielding against cosmic ray background, with about 70~muons per day passing through the detector.
The detector contains \SI{1}{kt} of heavy water, D$_2$O, in a spherical acrylic vessel with a diameter of \SI{12}{m}. This vessel is surrounded by a support structure with a diameter of \SI{17.8}{m}, which carries 9438 inward-looking photomultiplier tubes (PMTs) and is filled with purified light water, which is used as an active shield~\cite{Boger2000}.

The SNO detector is able to detect solar $^8$B neutrinos through three different reactions,
\begin{align}
\nue + d	&\rightarrow p + p + e^-	\tag{CC}\label{eq-nu-cc}\\
\nu + d	&\rightarrow p + n + \nu	\tag{NC}\label{eq-nu-nc}\\
\nu + e^-\!	&\rightarrow \nu + e^-\!,	\tag{ES}\label{eq-nu-es}
\end{align}
where $\nu$ refers to a neutrino with any one of the three flavors. This enables it to measure both the total neutrino flux (using elastic scattering~\eqref{eq-nu-es} and neutral current~\eqref{eq-nu-nc} interactions) and the electron neutrino flux (using charged current~\eqref{eq-nu-cc} interactions) in one experiment. The measured electron neutrino flux was lower than predicted by theory, consistent with prior experiments. The total neutrino flux, on the other hand, was found to be in agreement with the standard solar model prediction, thus proving that the solar neutrino flux contained multiple neutrino flavors~\cite{Ahmad2002}.

Since this direct experimental confirmation of neutrino oscillations, many experiments have tried to measure the four parameters of the PMNS matrix.

Even before the direct detection of neutrino oscillations by SNO, the Super-Kamiokande collaboration (see chapter~\ref{ch-nu-sk}) found an asymmetry in the flux of $\nu_\mu$ in particle showers produced in the Earth’s atmosphere by cosmic rays. The number of downward-going \numu\ events was significantly lower than that of upward-going \numu\ events. If this is interpreted as $\numu \leftrightarrow \nu_\tau$ oscillations of the upward-going neutrinos while they travel through the Earth, it can be used to determine the oscillation angle $\theta_{23}$~\cite{Fukuda1998}.

The second angle, $\theta_{12}$, was first determined in 2002 by SNO~\cite{Ahmad2002a} and in the following year through a combination of solar neutrino data and \nuebar\ disappearance measurements by the KamLAND experiment~\cite{Eguchi2003,Araki2005}.

Until recently, the third angle was still compatible with zero. Measurements that showed a first hint of a non-zero $\theta_{13}$ were announced in December 2011 by the Double Chooz collaboration~\cite{Abe2012}. Three months later, the Daya Bay experiment published results that showed $\theta_{13}$ to be significantly above zero~\cite{An2012}, which was confirmed just weeks later by the RENO collaboration~\cite{Ahn2012}.

Today, the best values for the mixing angles according to the Particle Data Group~\cite{PDG2014} are
\begin{align*}
\sin^2{2 \theta_{12}} &= 0.846 \pm 0.021\\
\sin^2{2 \theta_{13}} &= 0.093 \pm 0.008\\
\sin^2{2 \theta_{23}} &= 0.999 ^{+ 0.001} _{- 0.018},
\end{align*}
while the value of the CP-violating phase $\delta$ is still unknown~\cite{Capozzi2014}.

\subsection{The MSW Effect}\label{ch-nu-msw}
In addition to the vacuum oscillations described before, neutrino oscillations can be impacted by the presence of matter. Wolfenstein in 1978 pointed out that neutrino oscillations may be enhanced by coherent forward scattering on matter. While neutral-current scattering on neutrons is the same for all neutrino flavors and simply yields a common phase shift of little physical importance, charged-current scattering on electrons discriminates between \nue\ and other flavors~\cite{Wolfenstein1978}. Mikheev and Smirnov in 1985 studied the effect of a varying matter density on neutrino oscillations and applied their findings to solar neutrinos~\cite{Mikheev1985}. The effect was accordingly named the Mikheev-Smirnov-Wolfenstein (MSW) effect.
Shortly afterwards, Bethe gave a different derivation of the phenomenon described by Mikheev and Smirnov and applied it to the solar neutrino experiments by Davis and others~\cite{Bethe1986}.

For simplicity, we will describe this effect in a two neutrino flavor picture. Furthermore, we will disregard the influence of neutral-current scattering on neutrons, as discussed above. In the two flavor case, the relationship between flavor eigenstates and mass eigenstates is given by
\begin{equation}
\begin{pmatrix}
	\nue\\
	\nu_\mu
\end{pmatrix}
=
\begin{pmatrix}
	\cos{\theta_C} & \sin{\theta_C}\\
	- \sin{\theta_C} & \cos{\theta_C}
\end{pmatrix}
\begin{pmatrix}
	\nu_1\\
	\nu_2
\end{pmatrix}.
\end{equation}

\begin{figure}[ht]
	\centering
	\includegraphics[scale=0.45]{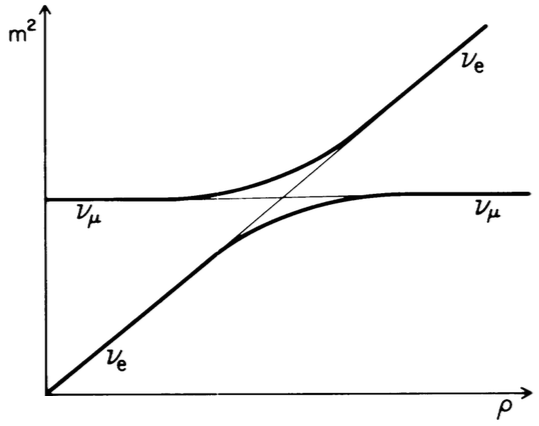}
	\caption{Impact of the presence of matter with density $\rho$ on the neutrino eigenstates. The thin lines show the effective masses of the flavor eigenstates \nue\ (diagonal) and $\nu_\mu$ (horizontal), while the thick lines show the evolution of the propagation eigenstates. As discussed in the text, we use a simplified two-flavor picture and ignore the effect of neutral-current scattering on neutrons, which is common to both flavors. (Figure taken from reference~\cite{Bethe1986}.)}
	\label{fig-nu-msw}
\end{figure}

Charged-current scattering on electrons in matter creates an effective potential $V_\text{eff} = \sqrt{2} G_F n_e$ for \nue\ (but not for $\nu_\mu$). This changes the effective squared mass of \nue\ to grow linearly with the electron number density $n_e$, which in the Sun and the Earth is proportional to the total matter density $\rho$.
In vacuum, the propagation eigenstates --~i.\,e. the eigenstates of the Hamiltonian~-- are equal to the mass eigenstates. In matter, the additional term $V_\text{eff}$ modifies the propagation eigenstates as shown in figure~\ref{fig-nu-msw}.

We note that the masses of propagation eigenstates do not cross over~-- contrary to those of the flavor eigenstates. Accordingly, when an electron neutrino is generated in the center of the Sun at a high density and then travels outwards, it propa\-gates through a density gradient following the upper thick line in figure~\ref{fig-nu-msw}. Throughout, it stays in the propagation eigenstate and is thus converted into a muon neutrino.

While the density inside the Earth is much smaller than in the center of the Sun, the matter will still cause the propagation eigenstates and mass eigenstates to split up and thus cause neutrino oscillations. This Earth matter effect and its detectability in supernova neutrino spectra has been the subject of intense research~\cite{Dighe2000,Lunardini2001,Dighe2003,Dighe2003a}. A more recent analysis, however, found that this effect is unlikely to be detectable in current or next-generation detectors~\cite{Borriello2012}.

The phenomenology of supernova neutrino oscillations is very rich and includes a number of additional effects. First calculations of the complete oscillation picture were published in recent years~\cite{Lund2013}, but the picture is still not clear and depends on properties of the progenitor.
Following reference~\cite{Tamborra2014}, we will consider both the case of no flavor swap (where all \nuebar\ detected on Earth are created as \nuebar\ at the supernova) and the case of full flavor swap (where all \nuebar\ detected on Earth are created as $\bar{\nu}_\mu$ or $\bar{\nu}_\tau$). We expect the actual results to be somewhere in between these two extremes.


\section{Neutrino Detection in Water Cherenkov Detectors}
There are several different types of neutrino detectors, which rely on different detection mechanisms. We will focus on water Cherenkov detectors, leaving aside liquid scintillator detectors (as used by Reines and Cowan~\cite{Cowan1953} or, more recently, by the KamLAND~\cite{Suzuki1999} and Borexino~\cite{Ranucci2001} experiments), radiochemical detectors (as used by Davis~\cite{Davis1955,Davis1968} or, more recently, by the GALLEX/GNO experiment~\cite{Kirsten2008} and SAGE~\cite{Abdurashitov2002}) and all other detector types that are currently in use.

In water Cherenkov detectors, inverse beta decay (IBD) is generally the main detection channel for supernova neutrinos, making up roughly 90\,\% of the events. Therefore, instead of calculating all detection channels in great detail for each detector, we will simply scale the IBD rate by the appropriate factor to take the other detection channels into account.

\subsection{Detection Mechanism}
\subsubsection{Inverse Beta Decay}
Water Cherenkov detectors use the inverse beta decay (IBD) reaction on hydrogen nuclei,
\begin{equation*}
\nuebar + p \rightarrow n + e^+,
\end{equation*}
to detect electron antineutrinos.

The threshold energy for this reaction, given by the difference between the proton mass and the sum of neutron and positron mass, is $E_{thr} \approx \SI{1.8}{MeV}$. In the case of much higher neutrino energy --~typically \SIrange{5}{50}{MeV} for supernova neutrinos~\mbox{--,} the resulting positron is highly relativistic. In this case, since its speed is higher than the speed of light in water ($c_{\text{H}_2 \text{O}} = c_\text{vac} / n_{\text{H}_2 \text{O}} \approx c_\text{vac} / 1.33$), it will radiate Cherenkov photons~\cite{Cherenkov1937}.
These photons will be radiated with an angle $\cos{\theta} = 1 / n\beta$, so in the high-energy limit, where $\beta \approx 1$, this angle is given by $\theta \approx \cos^{-1}(1/n) \approx 41^\circ$.

These Cherenkov photons are detected by photo multiplier tubes (PMTs) and the measured data is then transferred to a computing facility, where the measurements from multiple PMTs are combined.

\subsubsection{Neutron Detection}
The free neutron, which was produced in IBD, will scatter on nuclei in the medium and will finally get absorbed by a nucleus within seconds. The binding energy will be released in the form of photons which can also be detected by the PMTs.

\enlargethispage{-\baselineskip} 
As pointed out by Beacom and Vagins in 2004~\cite{Beacom2004}, the absorption time can be reduced to a few tens of microseconds by adding small amounts of gadolinium chloride (GdCl$_3$), whose thermal neutron capture cross section is about five orders of magnitude larger than those of hydrogen nuclei, to the detector material. The neutron capture on gadolinium would lead to a gamma cascade of \SI{8}{MeV}, which can be detected by the same PMTs as the Cherenkov photons.

This coincident detection of Cherenkov photons and the \SI{8}{MeV} gamma cascade from neutron capture is very unlikely to be imitated by background events and could thus guarantee a near-instant high-certainty detection of neutrino-induced IBD. It could also be used to improve directional information from supernova neutrinos, or to reduce backgrounds in the search for the diffuse supernova neutrino background~\cite{Beacom2004}. A test facility called EGADS (Evaluating Gadolinium’s Action on Detector Systems) was built starting in 2009~\cite{Adams2013}.

\subsubsection{Time Resolution}
The time resolution of the detector depends on the time resolution of the PMTs, which is typically within a few nanoseconds, and the spatial resolution of the event reconstruction. The latter is necessarily smaller than the detector’s total size, and can thus be at most a few tens of meters, corresponding to roughly \SI{100}{ns} (see figure~\ref{fig-nu-sk-event}). The total time resolution of the detector will therefore be smaller than \SI{1}{\micro s}.

The neutrinos from a supernova are highly relativistic, moving at speeds $v_\nu \approx c$. The time-of-flight delay caused by finite neutrino masses is~\cite{Lund2010}
\begin{equation}
\Delta t = \SI{0.57}{ms} \left( \frac{m_\nu}{\si{eV}} \right)^2 \left(\frac{ \SI{30}{MeV} }{E}\right)^2 \left(\frac{D}{ \SI{10}{kpc} }\right)^2,
\end{equation}
where $m_\nu$ is the mass of a neutrino species, $E$ is the neutrino energy and $D$ is the distance between the supernova and a detector.
For a galactic supernova with a typical distance of less than \SI{20}{kpc} (see figure~\ref{fig-or-sn-distances}), neutrino energies of \SIrange{5}{50}{MeV}, and a neutrino mass below \SI{0.2}{eV}\footnote{An upper limit of \SI{0.2}{eV} on the sum of the three neutrino masses can be derived from cosmological arguments~\cite{Hannestad2010}; the current experimental limit from $\beta$ decay of tritium is about one order of magnitude higher~\cite{Aseev2011,Kraus2005}. The KATRIN experiment, which is currently under construction, is expected to start data taking in late 2015 and to be sensitive to an electron neutrino mass of down to \SI{0.2}{eV} within five years~\cite{Parno2013}.},
this delay will typically be below \SI{1}{ms}.

Neutrino flux variations at time scales of greater than \SI{1}{ms} are therefore conserved on the way between the supernova and the detector. Time-dependent variations in the neutrino flux with a frequency of up to several \SI{100}{Hz} can potentially be observed in the detector signal.
\enlargethispage{-\baselineskip} 

\subsection{DUMAND and AMANDA}
Starting in the mid-1970s, astroparticle physicists discussed using sea water to build a large volume water Cherenkov detector under the name DUMAND (Deep Underwater Muon and Neutrino Detector)~\cite{Blood1976}.
The DUMAND design uses the ocean water in several different ways simultaneously: as a shield against low-energy cosmic rays, as a target for neutrino and muon interactions, and as a detection medium in which highly relativistic particles produce Cherenkov photons.
After more than a decade of research, construction of a first prototype began in the late 1980s.

The goal of DUMAND was to build a detector with a very large detection volume, in order to be able to observe muons or neutrinos with an energy of at least several \si{TeV}. The proposed setup consisted of 36 strings, each equipped with 21 equidistant optical modules, located off the coast of Hawaii at about \SIrange{4}{5}{km} below the ocean surface. The total instrumented volume was proposed to be $\SI{250}{m} \times \SI{250}{m} \times \SI{500}{m}$ or more than 30 megatons of water~\cite{Grieder1986}.

At a depth of \SIrange{4}{5}{km} below the ocean surface, the background from sunlight and from cosmic ray muons will be negligible. The main background will come from bioluminiscence and from the radioactive potassium isotope $^{40}$K, each responsible for an event rate of approximately \SI{100}{kHz} per optical module.
This background can be reduced significantly by looking for events that produce a signal in multiple optical modules within a very brief time interval. Such events are commonly generated by high-energy particles, which produce a shower of Cherenkov photons, but are unlikely to be generated by multiple independent radioactive decays. This coincident detection technique reduces the background rate at the cost of excluding events from lower-energy particles.
Due to the relatively coarse instrumentation, the detector would usually pick up only one Cherenkov photon from a supernova neutrino induced IBD. To the detector, this would be indistinguishable from background.

In the wake of the supernova 1987A, Pryor and others suggested small modifications to lower the energy threshold of DUMAND, which would have enabled it to detect the neutrino burst from a galactic supernova~\cite{Pryor1988}.
Several years later, Halzen and others noted that even the original DUMAND design could measure supernova neutrinos. They pointed out that, while any single event may be indistinguishable from background, the temporary increase in event rate across all optical modules could be statistically significant~\cite{Halzen1994}.

Unfortunately, DUMAND was cancelled after experiencing technical difficulties in the first deployment~\cite{Halzen2008}. Lessons learned from DUMAND were essential for later underwater neutrino detectors like the Baikal telescope in Lake Baikal~\cite{Belolaptikov1997} and ANTARES in the Mediterranean Sea~\cite{Ageron2011}.

Meanwhile, other groups worked on a detector called AMANDA (Antarctic Muon And Neutrino Detector Array). While the basic detector design --~a coarse grid of PMTs which uses water as shielding, detector material and propagation medium for Cherenkov photons~-- was similar to DUMAND, AMANDA was located at the South Pole and used a region of a \SI{3}{km} thick ice sheet as the detector material, instead of ocean water.

Compared to ocean water, ice is a very sterile medium. Bioluminiscence and radioactive decays of naturally occuring $^{40}$K, which are the predominant sources of background noise in ocean water, are negligible in the ice. The main sources of noise in AMANDA are the dark noise from the PMTs and traces of $^{40}$K in the glass housings of the optical modules. The total background rate is roughly \SI{1}{kHz} per optical module~\cite{Andres2000}~-- two orders of magnitude lower than in DUMAND. Additionally, it is possible to operate a surface detector on top of AMANDA, whose data can be used as a veto against air showers and for calibration.

Explorations for AMANDA began in 1991, and in 1993, the first strings were deployed to a depth of \SIrange{810}{1000}{m}. It was soon discovered, however, that the high concentration of air bubbles at this depth caused strong light scattering, which smeared out the arrival time of Cherenkov photons and made precise track reconstruction impossible. Starting in 1995, four strings were deployed to a depth of \SIrange{1545}{1978}{m}, where the phase transition from air bubbles into air-hydrate crystals has completed. While scattering was still nearly an order of magnitude stronger than in water, it was low enough to enable track reconstruction~\cite{Andres2000}.

Halzen and others had pointed out in 1994 that a supernova might be detectable in AMANDA because of a temporary increase of the event rate across all optical modules~\cite{Halzen1994} and later concluded that AMANDA could view a radius of \SI{17}{kpc}, covering most stars in our galaxy~\cite{Halzen1996}. Using the 1997 AMANDA setup consisting of ten strings and 302 optical modules, Ahrens and others found that AMANDA is able to cover about two thirds of the progenitor stars in our galaxy~\cite{Ahrens2002}.

\subsection{IceCube}\label{ch-nu-ic}
While construction of AMANDA was ongoing, planning for a successor, the IceCube detector, began. It shared the fundamental design as well as the location at the South Pole.

Built between 2004 and 2010, IceCube consists of 78 strings in a triangular grid pattern with a horizontal spacing of \SI{125}{m}. Each string is instrumented with 60 digital optical modules (DOMs) with a horizontal spacing of \SI{17}{m}. The DOMs are glass spheres containing a PMT and readout electronics, which digitize the signal and add a time stamp.
The total instrumented volume is approximately \SI{1}{km^3} of ice.

\begin{figure}[htbp]
\centering
\begin{minipage}[l]{6.9cm}
	\includegraphics[width=6.6cm]{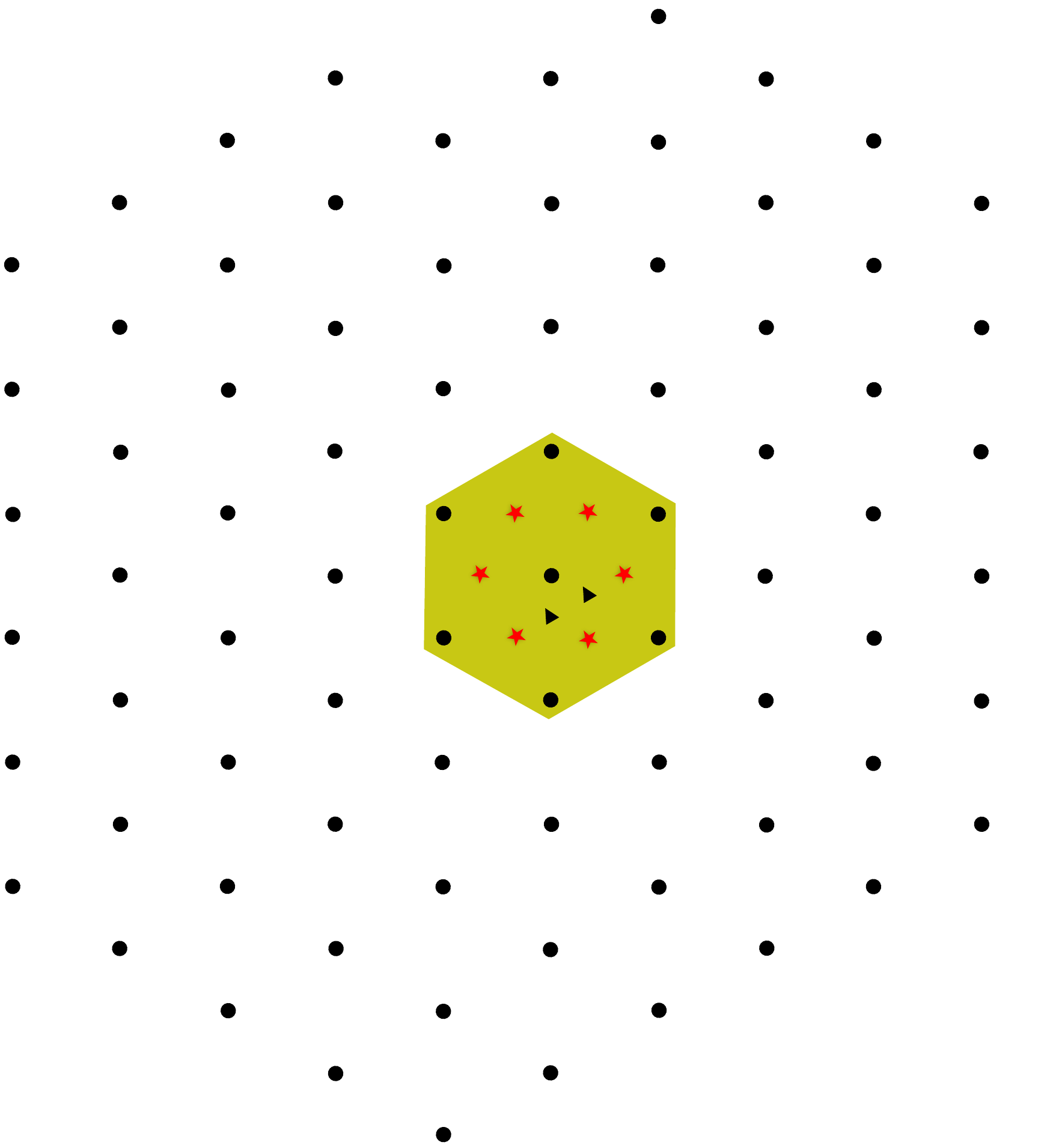}
\end{minipage}
\begin{minipage}[r]{6.9cm}
	\includegraphics[width=6.8cm]{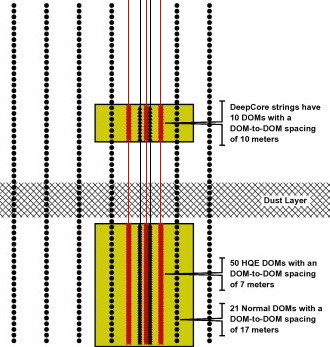}
\end{minipage}
\caption{Diagram showing the IceCube detector from the top (left) and the side (right), with DeepCore highlighted. (Side view taken from reference~\cite{Abbasi2012}.)}
\label{fig-nu-ic}
\end{figure}
Starting in 2009, eight additional strings instrumented with 60 DOMs each were added to create the DeepCore subdetector. Six of these strings (marked red in figure~\ref{fig-nu-ic}) have a horizontal string-to-string spacing of \SI{72}{m} and use PMTs with an improved “super bialkali” photocathode that improves optical sensitivity by about 35\,\%. The final two infill strings have an even smaller horizontal distance of \SI{42}{m} and use a mixture of both PMT models. All additional DeepCore strings have a smaller vertical distance of \SIrange{7}{10}{m} between DOMs~\cite{Abbasi2012}.

IceCube is designed to search for sources of neutrinos and cosmic rays with energies in the \si{TeV} range or higher.
The large instrumented volume is important for accurately measuring the energy of high energy particle showers, which requires that a large part of the particle shower is contained in the detector volume.
DeepCore increases the instrumentation density and thus lowers the energy threshold to the multi-\si{GeV} range.

For detection of supernova neutrinos, whose energy is typically below \SI{50}{MeV}, the DOM density in IceCube is very low. The effective volume for the detection of two photons from one IBD event is about two orders of magnitude smaller than the effective volume for detection of just one photon~\cite{Salathe2012}. Therefore, from any given IBD event, at most one photon will likely be detected.
Like DUMAND and AMANDA, IceCube will therefore detect the neutrino burst associated with a supernova as a temporary increase in the event rate across all DOMs. Since IceCube is unable to identify whether any given event during such a neutrino burst is caused by a supernova neutrino or not, it cannot provide event-by-event energy information.

Like AMANDA, IceCube benefits from the very sterile detector medium. In addition, IceCube uses detector components with reduced radioactivity, which leaves dark noise from the PMTs as the main background. The average background rate is \SI{540}{Hz} per DOM~\cite{Abbasi2011}.

The IBD rate per DOM in IceCube is given by~\cite{Abbasi2011}
\begin{equation}
R_\text{IBD, DOM} = N_\text{p, IC} \int \d E_e\, N_\gamma (E_e) \int \d E_\nu\, \tdiff{\Phi_\nu (E_\nu)}{E_\nu} \tdiff{\sigma (E_e, E_\nu)}{E_e},\label{eq-sn-ic-ibdom}
\end{equation}
where $\tdiffx{\sigma}{E_e}$ is the IBD cross section, differentiated with respect to the positron energy (see appendix~\ref{ap-IBD-cross-section}),  $\tdiffx{\Phi_\nu}{E_\nu}$ is the energy-dependent \nuebar\ flux (see appendix~\ref{ap-nu-flux}), $N_\gamma \approx 178 E_e / \si{MeV}$ is the number of Cherenkov photons produced by a positron with energy $E_e$, and $N_\text{p, IC}$ is the number of hydrogen nuclei, i.\,e. possible targets. In IceCube, this is given by $N_\text{p, IC} = n_\text{p} V_\text{eff}$, where $n_\text{p} = \SI{6.18e22}{cm^{-3}}$ is the number density of protons in ice ($\rho_\text{ice} = \SI{0.924}{g cm^{-3}}$) and $V_\text{eff}$ is the effective detection volume per DOM.

IBD is responsible for approximately $94\,\%$ of the neutrino events in IceCube~\cite{Abbasi2011}. (The other events are mostly from scattering on electrons and from electron neutrino and antineutrino capture on oxygen nuclei.) As discussed earlier, we can appro\-xi\-mate the total event rate per DOM by dividing the IBD rate per DOM by this scaling factor:
\begin{equation}\label{eq-nu-ic-om-rate}
R_\text{DOM} = \frac{R_\text{IBD, DOM}}{0.94}
\end{equation}

Additionally, every DOM shows a background rate of $r_\text{BG} \approx \SI{540}{s^{-1}}$. By introducing a dead time of $t_\text{dead} = \SI{250}{\micro s}$ after every hit, this background is reduced by nearly 50\,\% to $r_\text{BG} \approx \SI{286}{s^{-1}}$, while reducing detector uptime and thus the measured rate by approximately 13\,\%.
\footnote{In an earlier design of the IceCube data acquisition system, this dead time was enforced by a field programmable gate array in the DOM. In that design, a minimum dead time of \SI{110}{\micro s} was actually necessary to avoid an overflow of the 4-bit event counter in each \SI{1.6384}{ms} bin~\cite{Abbasi2011}. In situations where the event rate reaches $\sim 1 / t_\text{dead}$ --~for example, a supernova at $\lesssim \SI{2}{kpc}$~--, this could cause a “whiteout” problem, where the detector is flooded with photons and becomes unable to produce differentiated measurements beyond a maximum event rate.

This disadvantage was acknowledged by the IceCube collaboration and has since been eliminated in a recent extension to the data acquisition system. This was done by introducing a data buffer which stores every single hit at a $\mathcal{O}(\SI{10}{ns})$ time resolution~\cite{Baum2014}. 
For very close supernovae, it is thus possible to measure $R_\text{IC} = N_\text{DOM} R_\text{DOM}$ instead of equation~\eqref{eq-nu-ic-dead-time}. Since our work focusses on improving detection prospects in situations with low signal, we will continue to use the dead time technique in this thesis.}
Using this, the total signal rate in the IceCube detector is given by
\begin{equation}\label{eq-nu-ic-dead-time}
R_\text{IC} = N_\text{DOM} \frac{0.87 R_\text{DOM}}{1+ R_\text{DOM} t_\text{dead}},
\end{equation}
where $N_\text{DOM} = 5160$ is the number of DOMs in IceCube~\cite{Abbasi2011, Tamborra2014}.

\subsection{Super-Kamiokande}\label{ch-nu-sk}
The original Kamiokande (an abbreviation of “Kamioka Nucleon Decay Experiment”) detector~\cite{Nakamura1989} was a cylindrical tank containing \SI{3}{kt} of water, located underground in the Kamioka mine in the Gifu region in Japan. It started operations in 1983 looking for nucleon decay and has increased lower limits on the proton lifetime for a number of different decay channels.

Following detector upgrades, it produced a number of additional physics results, including observation of solar $^8$B neutrinos and observation of atmospheric neu\-tri\-nos.
In February 1987, Kamiokande detected a dozen neutrinos from SN1987A which until today remain half of all supernova neutrinos ever detected by humankind.

In 2002, part of the Nobel Prize for Physics was jointly awarded to Ray Davis Jr. and to the Kamiokande collaboration’s Masatoshi Koshiba for their “pioneering contributions to astrophysics, in particular for the detection of cosmic neutrinos”~\cite{NobelPrize2002}.

Kamiokande’s successor, the Super-Kamiokande detector~\cite{Fukuda2003}, started data-taking in 1996, after more than four years of construction. It is located in the Mozumi mine in Japan’s Gifu region below \SI{1000}{m} of rock shielding, corresponding to \num{2700}\,m.\,w.\,e.

It is a cylindrical tank, similar in form to its predecessor, with a diameter of \SI{39}{m} and a height of \SI{42}{m}, containing \SI{50}{kt} of water. This total volume is divided by a light-proof support structure into an outer shell with a width of approximately \SI{2.5}{m} and an inner region with a diameter of \SI{33.8}{m} and a height of \SI{36.2}{m}. The outer region is observed by photomultiplier tubes (PMTs) mounted on the outside of the support structure and acts as an active veto against incoming particles such as cosmic-ray muons and to shield against radioactivity from the surrounding rock.

The inner region contains roughly \SI{32}{kt} of water and includes a fiducial volume of approximately \SI{22.5}{kt} of water surrounded by a \SI{2}{m} wide shell to exclude ano\-ma\-lous events resulting from charged particles passing close to the PMTs, and to act as an additional shield against natural radioactive background from the surrounding rock.
The inner region was at first instrumented with \num{11146} inward-facing PMTs with a \SI{50}{cm} diameter, resulting in an effective photocathode coverage of 40\,\% of the inner detector’s surface. In November of 2001, during maintenance work in preparation for a detector upgrade, a chain reaction triggered by an imploding PMT destroyed more than half of the PMTs. The detector continued operations in 2002 with a reduced coverage of 19\,\% and was restored to full coverage for the third phase, starting in 2006~\cite{Abe2011a}.

A water purification system is employed to reduce scattering losses, which nega\-tive\-ly impact the detector’s energy resolution due to the delayed arrival time of the scattered photons at the PMTs.
For light at a wavelength of \SI{420}{nm}, which is close to the region of maximum efficiency of the PMTs, this results in an attenuation length of roughly \SI{100}{m}, about twice the diagonal size of the inner detector, which is close to \SI{50}{m}.

\begin{figure}[htp]
	\centering
	\includegraphics[scale=0.45]{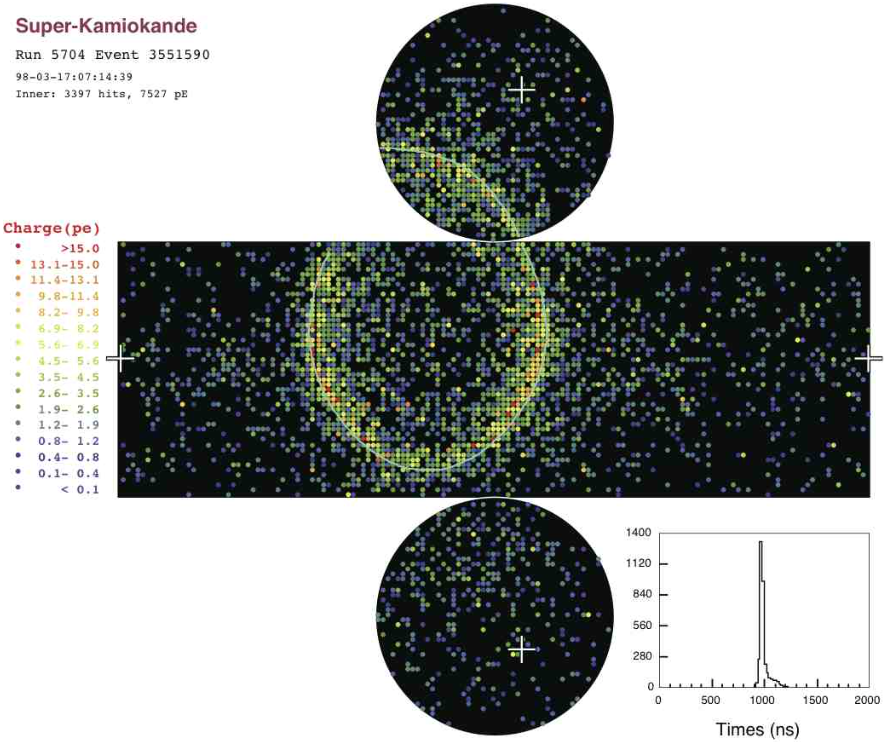}
	\caption{Unrolled view of the Super-Kamiokande detector with a Cherenkov ring from an electron-like event. Colors indicate the number of photo-electrons detected in each PMT. The plot in the lower right shows the distribution of arrival times, which has a sharp peak with a width of less then \SI{100}{ns}. (Figure taken from reference~\cite{Abe2011}.)}
	\label{fig-nu-sk-event}
\end{figure}

As discussed earlier, a highly relativistic positron produced in an IBD event sends out Cherenkov light in a cone with an angle of about 41$^\circ$.
While IceCube, due to its coarse instrumentation, is only able to detect at most one photon from any given IBD event, Super-Kamiokande can detect multiple photons from every IBD event. This produces an easily identifiable circle of hits in the PMTs (see figure~\ref{fig-nu-sk-event}) and allows a very accurate reconstruction of events with neutrino energies as low as \SI{4.5}{MeV}~\cite{Abe2011a}.

Coincident detection techniques can be used to distinguish between multiple events, and can also be used to identify and suppress uncorrelated noise like dark noise of the PMT or radioactive decays in the PMT material. Other background from radioactive decays in the detector medium or from penetrating cosmic rays may be identifiable due to its reconstructed energy. As a result, the background in Super-Kamiokande is very small and we will ignore it in the remainder of this thesis.

For Super-Kamiokande, the IBD rate is given by
\begin{equation}\label{eq-nu-sk}
R_\text{IBD, SK} = N_\text{p, SK} \int \d E_e \int \d E_\nu\, \tdiff{\Phi_\nu (E_\nu)}{E_\nu} \tdiff{\sigma (E_e, E_\nu)}{E_e}.
\end{equation}
The number of hydrogen nuclei is $N_\text{p, SK} \approx \num{2.1e33}$ in the inner detector region and $N_\text{p, SK} \approx \num{1.5e33}$ in the fiducial volume.

\subsection{Hyper-Kamiokande}\label{ch-nu-hk}
Hyper-Kamiokande~\cite{Abe2011} is the proposed successor to Super-Kamiokande. If construction, which is expected to start in 2018, goes according to schedule, data-taking could start in the year 2025~\cite{Normile2015}.
Its proposed location is about \SI{8}{km} south of Super-Kamiokande at \SI{648}{m} underground, corresponding to a shielding of \num{1750}~m.\,w.\,e.

\begin{figure}[ht]
	\centering
	\includegraphics[scale=0.32]{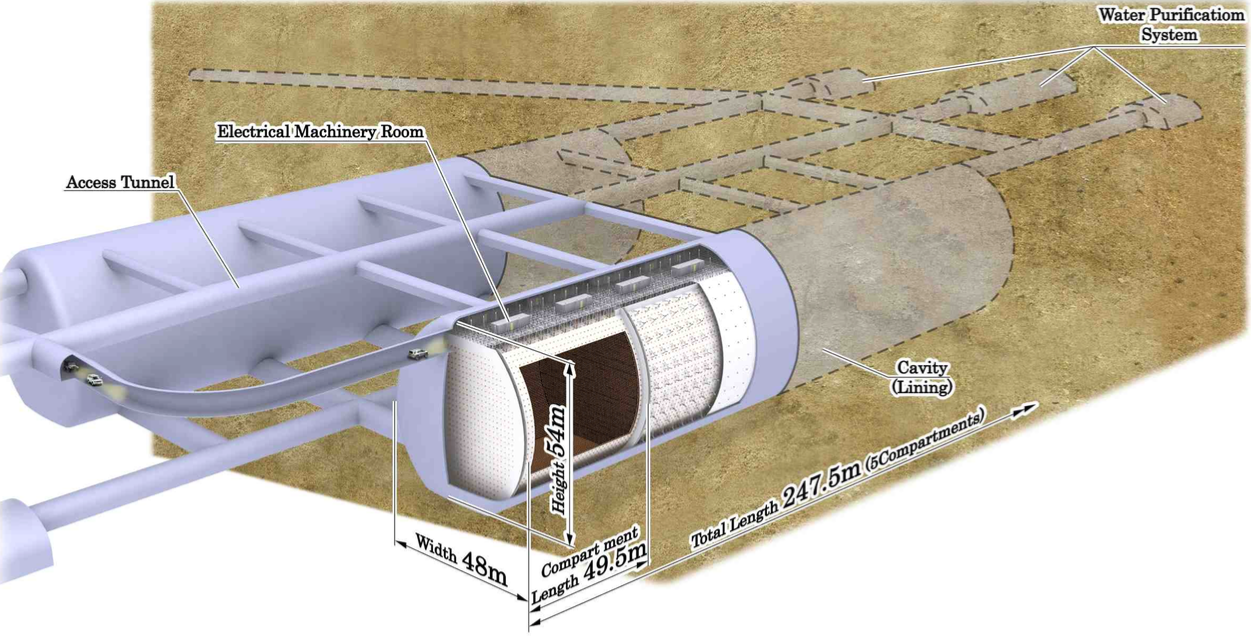}
	\caption{Schematic view of the Hyper-Kamiokande detector. (Figure taken from reference~\cite{Abe2011}.)}
	\label{fig-nu-hk}
\end{figure}
\enlargethispage{-\baselineskip} 

Hyper-Kamiokande follows the same basic design as its predecessor. It consists of two nearly cylindrical tanks with outer dimensions of $\SI{48}{m} \times \SI{54}{m} \times \SI{247.5}{m}$ each. The total detector mass is $\SI{990}{kt}$ (about 20 times the size of Super-Kamiokande), containing an inner region of $\SI{740}{kt}$, which includes a fiducial volume of $\SI{560}{kt}$ (about 25 times the size of Super-Kamiokande’s fiducial volume). The surrounding outer region with a thickness of \SI{2}{m} will be viewed with \num{25000} PMTs in order to act as an active veto against cosmic-ray muon backgrounds and to shield against radioactivity from the surrounding rock.

The inner region of each of the two tanks will be split into five separate compartments with a length of \SI{49.5}{m} each, making each compartment comparable in size to the whole Super-Kamiokande detector. The inner region will be equipped with PMTs and is planned to reach a photocathode coverage of 20\,\%, about half the maximum coverage reached in the Super-Kamiokande detector. As a baseline, this goal could be reached by using \num{99000} PMTs of the type used in Super-Kamiokande with a \SI{50}{cm} diameter. Additionally, improvements are possible either through a reworked design of the protective shells for the PMTs or through use of a different PMT model with better performance characteristics.

The water purification system of the detector will be upscaled to a higher capacity, in order to reach the same water quality as in Super-Kamiokande. For light at a wavelength of about \SI{400}{nm}, this is expected to result in the same attenuation length of about \SI{100}{m}, still comfortably larger than the diagonal size of each compartment, which is roughly \SI{70}{m}.

Accordingly, Hyper-Kamiokande will also offer event-by-event energy information with an energy resolution that should be comparable to that of its predecessor.


Similar to its predecessor, Hyper-Kamiokande’s IBD rate is given by
\begin{equation}
R_\text{IBD, HK} = N_\text{p, HK} \int \d E_e \int \d E_\nu\, \tdiff{\Phi_\nu (E_\nu)}{E_\nu} \tdiff{\sigma (E_e, E_\nu)}{E_e}
\end{equation}
but the number of hydrogen nuclei increases to $N_\text{p, HK} \approx \num{3.7e34}$ for the fiducial volume and $N_\text{p, HK} \approx \num{4.9e34}$ for the whole inner detector region.

IBD is responsible for 88\,\% to 89\,\% of the neutrino events in Hyper-Kamiokande~\cite{Ikeda2007}\footnote{The cited paper gives a value of 88\,\% to 89\,\% for Super-Kamiokande. Given the strong similarities in the design of both detectors, we expect this to be applicable to Hyper-Kamiokande as well. Note, also, that even a deviation of several percentage points from this number would have very little influence on the results of this thesis, since we are concerned not with detectability \emph{per se}, but with the detectability in certain energy ranges \emph{relative to each other}.}, depending on the amount of flavor mixing. As discussed earlier, the total event rate can therefore be approximated by dividing the IBD rate by this scaling factor:
\begin{equation}\label{eq-nu-hk}
R_\text{HK} = \frac{R_\text{IBD, HK}}{0.89}
\end{equation}

While its detector volume will be three orders of magnitude smaller than that of IceCube, Hyper-Kamiokande’s much denser instrumentation means that it will reach about one third of the IceCube event rate in the case of a future supernova. Additionally, it features a background that is several orders of magnitude lower and for most purposes negligible. As pointed out by Tamborra and others, it will therefore be able to produce a better signal than IceCube for supernovae at large distances~\cite{Tamborra2013}. We will consider this in detail in chapter~\ref{ch-or-snr}.

Hyper-Kamiokande will also offer event-by-event energy information, which is not available in IceCube in the relevant energy range. This energy information has not been used in previous analyses of supernova neutrino detection. In chapter~\ref{ch-or-energy-dependence}, we will discuss whether this additional information can be used to improve the detection prospects of time-dependent variations in the neutrino number flux.

\chapter{Data Processing and Analysis}\label{ch-data}

We analyzed data from a simulation of a $27\ M_\odot$ progenitor by the Garching group~\cite{Hanke2013}. The raw data is available from the Garching Core-Collapse Supernova Archive\footnote{\url{http://www.mpa-garching.mpg.de/ccsnarchive/archive.html}}. The data set used in this thesis was preprocessed by Irene Tamborra; it is the same one that was used in reference~\cite{Tamborra2014}.

We will start this chapter by describing the simulation and the preprocessing of the raw data. Next, we will calculate the event rate in the IceCube detector including a shot noise contribution. We will then describe how we calculate the power spectrum of the event rate and use the power spectrum to identify the SASI oscillations in the event rate.

In chapter~\ref{ch-or}, we will apply the data analysis techniques described in this chapter to data from the Hyper-Kamiokande detector.
Throughout both chapters, we will show the data for two extreme cases --~no flavor swap (NFS; where all \nuebar\ detected on Earth were created by the supernova as \nuebar) and full flavor swap (FFS; where all \nuebar\ detected on Earth were created as \nux)~--, as discussed at the end of chapter~\ref{ch-nu-msw}.
In all calculations, we will assume a supernova at a fiducial distance of \SI{10}{kpc}.

\section{Preparation of the Data Set}\label{ch-data-dataset}

\subsection{Description of the Simulation}
All calculations in this thesis are based on data from a three-dimensional simulation by the Garching group~\cite{Hanke2013} for the $27\,M_\odot$ progenitor by Woosley and others~\cite{Woosley2002}.

The simulation was performed with the \textsc{Prometheus-Vertex} code, which combines the hydrodynamics solver \textsc{Prometheus} with the neutrino transport code \textsc{Vertex}~\cite{Rampp2002}. It included a detailed treatment of energy-dependent neutrino interactions and used a ray-by-ray-plus approach, which takes into account not only radial transport of neutrinos, but also non-radial neutrino advection and pressure terms. The simulation used Newtonian gravity but included an additional effective gravitational potential to account for modifications from General Relativity~\cite{Hanke2013}.

\subsection{Preprocessing and Selection of Observer Direction}
The raw data from the simulation contains the energy flux, mean energy and lumi\-nosity of the supernova neutrinos at a radius of \SI{500}{km} for different angular zones and for each of the flavors \nue, \nuebar\ and \nux, respectively.

This data was preprocessed as described in appendix~A of reference~\cite{Tamborra2014}.
The resulting data set contains the mean energy, mean squared energy and luminosity for \nue, \nuebar\ and \nux\ during the time from \SIrange{10.5}{552.1}{ms} after the core-bounce (phase 3 in chapter~\ref{ch-sn-phases}). Radial fluxes are given for three observer directions: “Violet” and “Light Blue” refer to opposite directions within the plane where SASI originates, which results in a SASI amplitude that is approximately maximal, while the “Black” direction lies outside of that plane and sees a smaller SASI amplitude.

We used the data for the observer direction with a maximum SASI amplitude, referred to as the  “Violet Direction” in reference~\cite{Tamborra2014}.
Choosing a different observer direction would simply result in a lower SASI amplitude, leading to a weaker signal and, therefore, a lower detection probability. We will not discuss other observer directions in this thesis, since we want to focus not on assessing detectability of SASI oscillations, but on new methods of data analysis which make use of the event-by-event energy information afforded by the Hyper-Kamiokande detector.

The preprocessed data set had a slightly varying time step-size of \SI{0.50 \pm 0.01}{ms}. In addition, there were a number of gaps (where adjacent data points had a time difference of roughly \SI{1}{ms}) or duplicates (where adjacent data points had a time difference of much less than \SI{0.5}{ms}).
To avoid any problems that might otherwise arise from these small irregularities, we rebinned the data set. To that end, we interpolated between the available data points and integrated the interpolated function in steps of the bin length, $t_\text{bin}$, to get the number of events per bin. In our tests, the influence of $t_\text{bin}$ on the conclusions has been negligible; therefore, we decided on $t_\text{bin} = \SI{1}{ms}$ mainly for convenience.

\section{Event Rate in IceCube}\label{ch-data-eventrate}

\begin{figure}[htbp]
	\centering
	\includegraphics[scale=0.48]{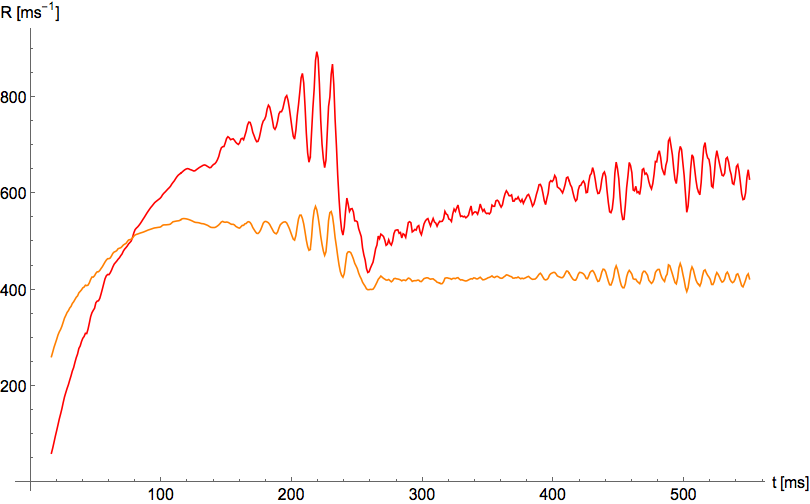}
	\caption{Total event rate in IceCube in the NFS (red) and FFS (orange) cases for a supernova at a fiducial distance of \SI{10}{kpc}.}
	\label{fig-data-total-rate}
\end{figure}

Using the data set described in the previous section, we can now use equation~\eqref{eq-nu-ic-dead-time} to calculate the expected event rate in the IceCube detector, $R_\text{IC}$. (See figure~\ref{fig-data-total-rate}.) We find that our results agree with those given by Tamborra and others in figure~15 of reference~\cite{Tamborra2014}. The SASI oscillations at \SIrange{200}{230}{ms} are clearly visible.
\begin{figure}[htbp]
	\centering
	\includegraphics[scale=0.48]{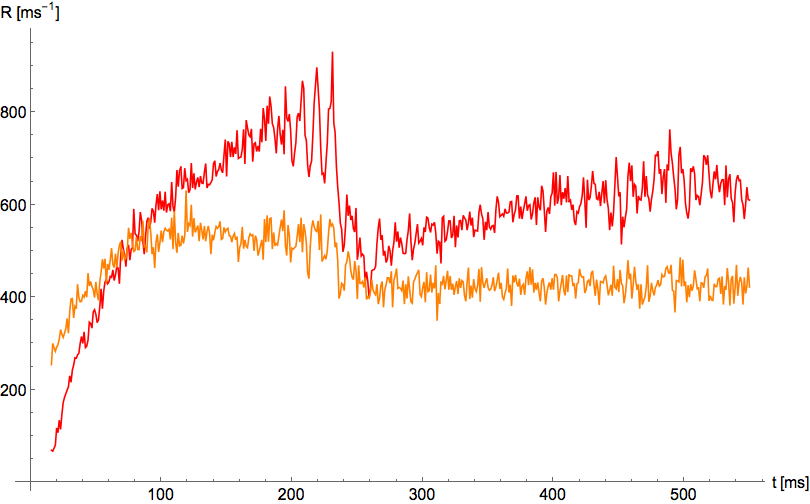}
	\caption{Same data as in figure~\ref{fig-data-total-rate}, but including a simulated shot noise.}
	\label{fig-data-poissoned-total-rate}
\end{figure}

In any realistic setting, however, the detected data will not look as smooth as in figure~\ref{fig-data-total-rate}. Instead, it will contain a shot noise contribution, which we simulate using a Poisson distribution for each \SI{1}{ms} bin.

For the shot noise, in addition to the supernova event rate, we have to take into account the background, which, at \SI{286}{s^{-1}} per optical module, is $R_\text{BG} \approx \SI{1476}{ms^{-1}}$ for the whole IceCube detector. The bin-to-bin fluctuations, which scale as $\sqrt{N}$ for a Poisson distribution with $N$ events, will therefore reach a minimum of $\sqrt{R_\text{BG}} \approx 38$ for \SI{1}{ms} bins, even if no supernova signal is present.
The shot noise is therefore given by $\sqrt{R_\text{BG} + R_\text{IC}}$.
One example for a resulting distribution can be seen in figure~\ref{fig-data-poissoned-total-rate}.

\section{Power Spectrum of the Total SASI Signal}\label{ch-data-power-spectrum}

For a more detailed analysis of the data, we will now calculate the power spectrum of the event rate. We follow the approach given in appendix~A of reference~\cite{Lund2010}, which we will describe in the remainder of this chapter.

\subsection{Fourier Transform}\label{ch-data-ft}
\begin{figure}[htb]
	\centering
	\includegraphics[scale=0.5]{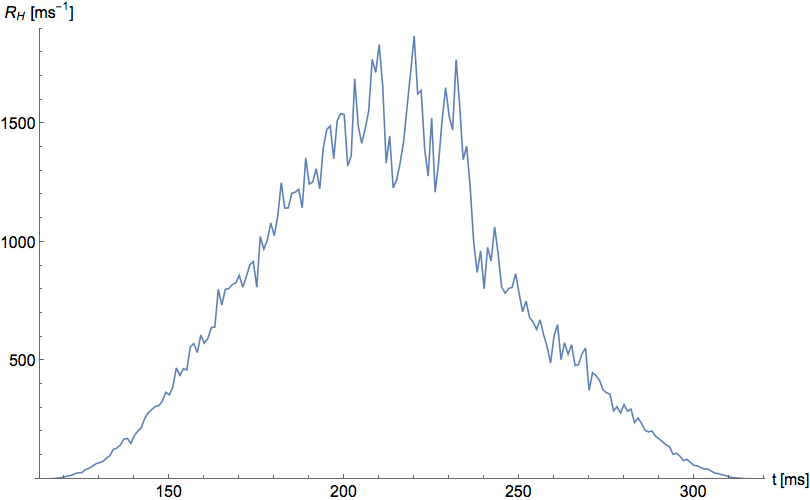}
	\caption{Total event rate in IceCube in the NFS case, multiplied by the Hann window function.}
	\label{fig-data-hanned-rate}
\end{figure}

We focus on the time interval of duration $\tau = \SI{200}{ms}$ from \SIrange{115}{315}{ms}. As we can see in figure~\ref{fig-data-total-rate}, this interval contains in its middle at about \SIrange{200}{230}{ms} the strongest SASI oscillations.
Next, we multiply the event rate in this time interval by the Hann window function
\begin{equation}
w(t) = 1 - \cos (2 \pi t / \tau) \label{eq-data-hann}
\end{equation}
to suppress edge effects in the Fourier transform.
This function is normalized such that its average weight is unity:
\begin{equation}
\int_0^\tau \d t\, \frac{w(t)}{\tau} = 1
\end{equation}
A plot of the resulting function, $R_H (t) = w(t) R(t)$, is displayed in figure~\ref{fig-data-hanned-rate}.
The Fourier transform of $R_H (t)$ is given by
\begin{equation}
h(f) = \int_0^\tau \d t\, R_H (t)\, e^{-i 2\pi f t}
\end{equation}
or, for the discrete case with equal bin width $\Delta$ and $t_j = j \Delta$,
\begin{equation}
h(f_k) = \Delta \sum_{j = 0}^{N_\text{bins} - 1} R_H (t_j)\, e^{-i 2 \pi f_k t_j}. \label{eq-data-h}
\end{equation}

\enlargethispage{-\baselineskip} 
We use a data set with the duration $\tau = \SI{200}{ms}$, which is split into $N_\text{bins} = 200$ bins, each of which has the width $\Delta = \SI{1}{ms}$. The Fourier transform of the binned data gives values for the discrete frequencies $f_k = k \cdot \delta f$, with the frequency spacing $\delta f = 1/ \tau = \SI{5}{Hz}$. The highest resolvable frequency (or “Nyquist frequency”) is $f_\text{max} = 1/ (2\Delta ) = \SI{500}{Hz}$.

For $f = 0$ and $f = f_\text{max}$, we define the spectral power as
\begin{equation}
P(0) = \frac{\abs{h(0)}^2}{N_\text{bins}^2} \quad \text{and}\quad
P(f_\text{max}) = \frac{\abs{h(f_\text{max})}^2}{N_\text{bins}^2},
\end{equation}
while we define it as
\begin{equation}
P(f_k) = \frac{\abs{h(f_k)}^2 + \abs{h(-f_k)}^2}{N_\text{bins}^2} = 2 \frac{\abs{h(f_k)}^2}{N_\text{bins}^2}
\end{equation}
for all other frequencies. The latter equality applies because the rate $R_H (t_j)$ in equation~\eqref{eq-data-h} is real, and therefore $\abs{h(f_k)} = \abs{h(- f_k)}$.
\enlargethispage{-\baselineskip} 

\subsection{Normalization of the Power Spectrum}\label{ch-data-normalization}

We will now normalize the power spectrum to the spectral power of the shot noise. In IceCube, detectability of fast signal variations for a supernova at the fiducial distance of \SI{10}{kpc} is limited by the shot noise of the background~\cite{Lund2010}. In Hyper-Kamiokande, on the other hand, the background is close to zero and the shot noise of the neutrino signal is the limiting factor.

Our aim is, therefore, to calculate the shot noise of the “equivalent flat” (e.\,f.) neutrino signal~-- that is, a neutrino signal that lacks the imprinted SASI oscillations, but is equivalent to the available signal in all other regards. Such an idealized signal is, of course, impossible to generate from the available data, let alone from first principles. Instead, as a reasonable approximation, we use a running average of the available signal.
Since the oscillations have a period of roughly \SI{12}{ms}, we will average over $\pm \SI{6}{ms}$.
Using this approximation, we calculate the power of the shot noise in appendix~\ref{ap-shot-noise}. We find that it does not depend on the specifics of the e.\,f. approximation.

\begin{figure}[tbp]
	\centering
	\includegraphics[scale=0.5]{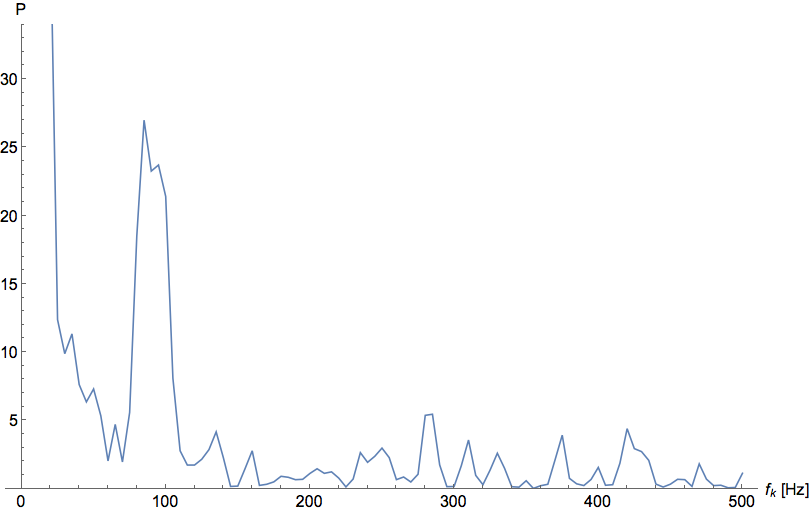}
	\caption{Power spectrum of the total SASI signal in IceCube for the NFS case.}
	\label{fig-data-power-spectrum}
\end{figure}

\begin{figure}[htbp]
	\centering
	\includegraphics[scale=0.5]{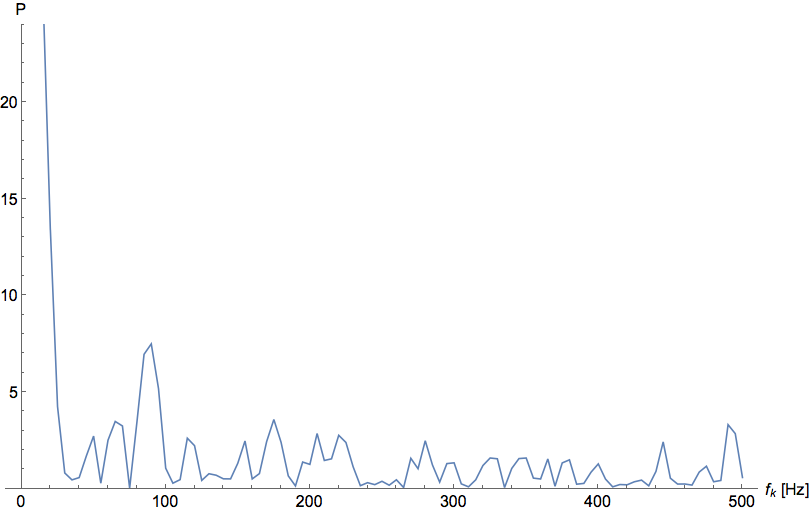}
	\caption{Power spectrum of the total SASI signal in IceCube for the FFS case.}
	\label{fig-data-power-spectrum-x}
\end{figure}

Having defined the Fourier transform (in the previous section), as well as the normalization of the power spectrum (in this section), we can now plot the power spectrum. The results are displayed in figures~\ref{fig-data-power-spectrum} and~\ref{fig-data-power-spectrum-x}.


We can easily discern two main features: the low frequency background at \mbox{$f \lesssim \SI{30}{Hz}$} and a peak at $f \approx \SI{80}{Hz}$, which corresponds to an oscillation with a period of roughly \SI{12}{ms}~-- exactly the SASI oscillations that are visible by eye in the signal in figure~\ref{fig-data-total-rate}!

As a simple example, we model the SASI oscillations as a sinusoidal wave with a frequency $f_\text{SASI}$; i.\,e. we assume that the signal is proportional to $1 + a\cdot \sin\left(2 \pi f_\text{SASI} t \right)$. The power spectrum of this signal has peaks at $f=0$ and $f = f_\text{SASI}$, with relative height given by
\begin{equation}
\frac{P(f_\text{SASI})}{P(0)} = \frac{a^2}{4}.
\end{equation}
In appendix~\ref{ap-shot-noise}, we calculated the average power of the shot noise, resulting in
\begin{equation}
\frac{\mean{P_{f≠0}}}{P(0)} = \frac{3}{2N}.
\end{equation}
For the sinusoidal modulations to be visible in the power spectrum, their power needs to be higher than that of the shot noise, $P(f_\text{SASI}) > \mean{P_{f≠0}}$, which requires an amplitude
\begin{equation}\label{eq-data-amin}
a > \sqrt{\frac{6}{N}}.
\end{equation}

In the time interval from \SIrange{100}{300}{ms}, the number of events $N$ from a supernova at the fiducial distance of \SI{10}{kpc} will be $\mathcal{O}(\num{e5})$ in both IceCube and Hyper-Kamiokande. This results in a minimum amplitude of about $1\,\%$. Of course, the power of the shot noise will often be higher than its average, so the amplitude will need to be at least a few percent in order to reliably identify the associated peak in the power spectrum above the shot noise background.


\chapter{Detectability of Fast Time Variations in Hyper-Kamiokande}\label{ch-or}

After introducing the data analysis techniques in the previous chapter, we will now apply them to the data available in the future Hyper-Kamiokande detector. As in the previous chapter, we will give results for the two extreme cases~-- no flavor swap (NFS) and full flavor swap (FFS).

At first, we will briefly discuss the relative detection prospects of the IceCube and Hyper-Kamiokande detectors depending on the supernova distance.

We will then use the event-by-event energy information that is available from Hyper-Kamiokande but not from IceCube. Assuming a supernova at the fiducial distance of \SI{10}{kpc}, we will investigate whether the amplitude of the SASI oscillation is a function of neutrino energy and whether the SASI detection probability in Hyper-Kamiokande can thus be improved by focussing on a suitable energy range.
Finally, we will artificially increase the amplitude of the oscillation of the mean energy and examine the results generated in this larger oscillation (LO) case.

For the most part, these calculations are equally valid for Super-Kamiokande. However, since its inner detector volume is about 25 times smaller than Hyper-Kamiokande’s, it would require a supernova at one fifth the distance to give the same statistics.

\section{Signal-to-Noise Ratio as a Function of Distance}\label{ch-or-snr}
We use equation~\eqref{eq-nu-hk} to calculate the event rate in the Hyper-Kamiokande detector, $R_\text{HK}$. As in chapter~\ref{ch-data-eventrate}, we include a shot noise contribution, which we simulated using a Poisson distribution for each \SI{1}{ms} bin. Since the background in the Hyper-Kamiokande detector is negligible, the shot noise is simply given by $\sqrt{R_\text{HK}}$.

\begin{figure}[htbp]
	\centering
	\includegraphics[scale=0.45]{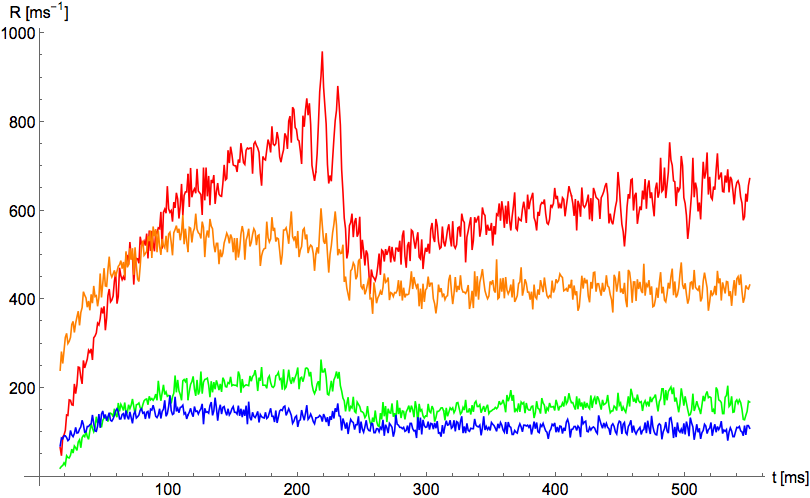}
	\caption{Total event rate in IceCube (NFS: red; FFS: orange) and Hyper-Kamiokande (NFS: green; FFS: blue) for a supernova at a fiducial distance of \SI{10}{kpc}, including a simulated shot noise.}
	\label{fig-or-poissoned-total-rate}
\end{figure}

In figure~\ref{fig-or-poissoned-total-rate}, we show the resulting event rate and include the IceCube event rate for comparison.
\begin{figure}[htbp]
	\centering
	\includegraphics[scale=0.45]{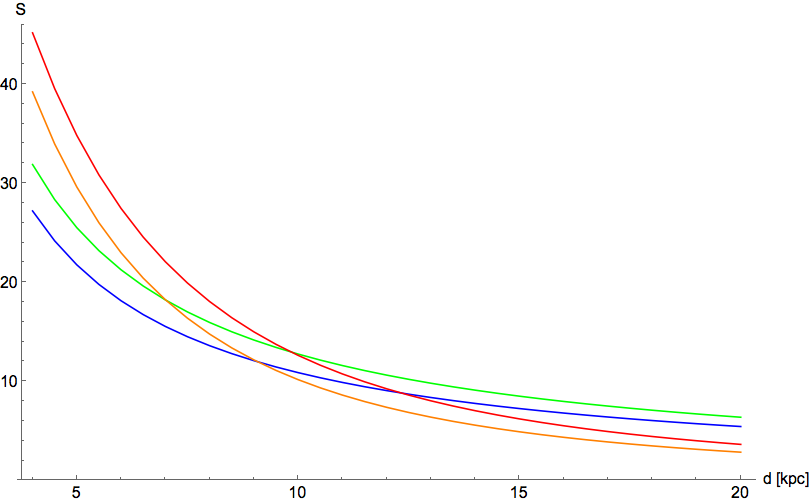}
	\caption{Signal-to-noise ratio $\mathcal{S}$ as a function of the supernova distance for IceCube (NFS: red; FFS: orange) and Hyper-Kamiokande (NFS: green; FFS: blue).}
	\label{fig-or-snr}
\end{figure}

As a first, very simple measure for the relative detection prospects of the IceCube and Hyper-Kamiokande detectors, we will calculate their signal-to-noise ratios as the quotients of their total event rate divided by their bin-to-bin fluctuations.

For IceCube, the shot noise is given in chapter~\ref{ch-data-eventrate}. The signal-to-noise ratio as a function of the supernova distance $d$ is
\begin{equation}
\mathcal{S}_\text{IC} (d) = \frac{ \mean{R_\text{IC}(d,t)} }{ \sqrt{R_\text{BG} + \mean{R_\text{IC}(d,t)}} }.
\end{equation}
(We write $\mean{R_\text{IC}(d,t)}$ to signify the average event rate per \SI{1}{ms} bin during the simulated time.)
For Hyper-Kamiokande, on the other hand, the background is negligible, and the bin-to-bin fluctuations are simply given by $\sqrt{R_\text{HK}(d,t)}$. The signal-to-noise ratio $\mathcal{S}_\text{HK} (d) = \sqrt{\mean{R_\text{HK} (d,t)}}$ is plotted in figure~\ref{fig-or-snr}.

For very close supernovae, the IceCube event rate $R_\text{IC}$  is much higher than the constant background. Accordingly, the shot noise in IceCube is $\sqrt{R_\text{BG} + R_\text{IC}} \approx \sqrt{R_\text{IC}}$ and at small distances the signal-to-noise ratio $\mathcal{S}$ is proportional to $\sqrt{R}$ for both detectors.

Figure~\ref{fig-or-snr} shows that IceCube’s signal-to-noise ratio is superior to Hyper-Kamiokande’s at distances below \SI{9}{kpc} for FFS and below \SI{10}{kpc} for NFS, while Hyper-Kamiokande is superior for larger distances.

\begin{figure}[htbp]
	\centering
	\includegraphics[scale=0.45]{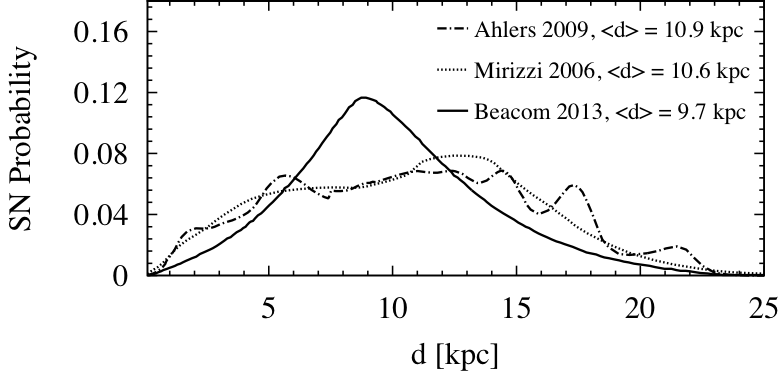}
	\caption{Estimates for the galactic supernova distribution as a function of distance from Earth~\cite{Mirizzi2006, Ahlers2009, Adams2013}. (Figure taken from~\cite{JUNO2015}.)}
	\label{fig-or-sn-distances}
\end{figure}

Figure~\ref{fig-or-sn-distances} shows three different estimates for the galactic supernova rate as a function of distance, as well as their average distance. The average distance is at around \SI{10}{kpc} and the distributions are roughly symmetric around that average, so Hyper-Kamiokande will have superior statistics for the next galactic supernova with a probability of about $0.5$.

\section{Energy Dependence of the SASI Amplitude}\label{ch-or-energy-dependence}

As a next step, we want to determine whether there is a phase or amplitude difference in SASI oscillations between different neutrino energies. If such a difference exists, the SASI oscillations in one energy range would be stronger than in the total data set. In this case, it could be beneficial to examine only a certain energy range instead of the total signal.

For the following calculations, we will assume that the distribution of neutrino energies can be approximated by a Gamma distribution $f(E, t)$ as discussed in appendix~\ref{ap-nu-flux}.
The event rate within an energy range $E_1 < E < E_2$ is given by
\begin{equation}
R (E_1, E_2, t) = R(t) \int_{E_1}^{E_2} \d E\, f(E, t).
\end{equation}

\subsection{No Flavor Swap}
\begin{figure}[htp]
	\centering
	\includegraphics[scale=0.5]{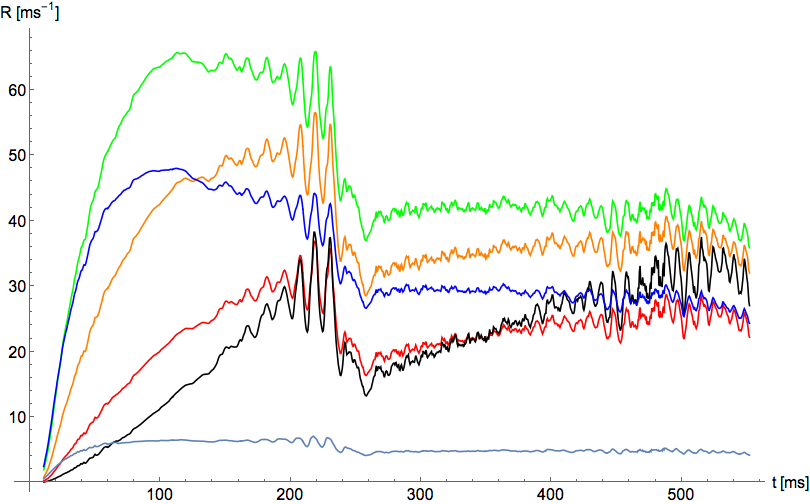}
	\caption{Comparison of event rates for different energy bins in the NFS case: $<$\SI{5}{MeV} (gray-blue), \SIrange{5}{10}{MeV} (blue), \SIrange{10}{15}{MeV} (green), \SIrange{15}{20}{MeV} (orange), \SIrange{20}{25}{MeV} (red), $>$\SI{25}{MeV} (black)}
	\label{fig-or-rate-5mev-bins}
\end{figure}

\enlargethispage{\baselineskip} 
Figure~\ref{fig-or-rate-5mev-bins} shows a comparison of this rate for a number of different energy bins. It is easy to observe that the SASI oscillations in all bins are in phase. Focussing on the relative amplitude, we notice some less obvious differences. To identify the magnitude of these differences, we display the six extrema in the range from \SIrange{200}{235}{ms} (the period with strongest SASI oscillations) in table~\ref{tab-or-extrema}.
Additionally, we display the relative amplitude
\begin{equation}\label{eq-or-arel}
a_\text{rel} = \frac{\mean{\text{max}} - \mean{\text{min}}}{\mean{\text{max}} + \mean{\text{min}}},
\end{equation}
where \mean{\text{max}} is the average of the three maxima ($2^\text{nd}$, $4^\text{th}$ and $6^\text{th}$ extremum), while \mean{\text{min}} is the average of the three minima ($1^\text{st}$, $3^\text{rd}$ and $5^\text{th}$ extremum),
and the relative signal-to-noise ratio
\begin{align}\label{eq-or-srel}
\mathcal{S}_\text{rel} (E_1, E_2) &= \frac{1}{a_\text{rel} (0, \infty) \sqrt{\mean{R(t)}}} \frac{a_\text{rel}(E_1, E_2)\, \mean{R(E_1, E_2, t)} }{ \sqrt{\mean{R(E_1, E_2, t)}} }\\
&= \frac{a_\text{rel} (E_1, E_2)}{a_\text{rel} (0, \infty)} \sqrt{\int_{E_1}^{E_2} \d E_\nu\, f(E_\nu)},
\end{align}
which is the ratio of the absolute amplitude of the SASI oscillations to the bin-to-bin fluctuations discussed in section~\ref{ch-or-snr}. For ease of comparison, $\mathcal{S}_\text{rel}$ is normalized such that the value for the total signal is $\mathcal{S}_\text{rel} (0, \infty) = 1$. The detection prospects in the energy bin between $E_1$ and $E_2$ are better than in the total signal if $\mathcal{S}_\text{rel} (E_1, E_2) > 1$.
In appendix~\ref{ap-srel-derive}, we show that this condition can also be derived by considering the spectral power of a signal.

\begin{table}[htbp]
	\begin{center}
	\begin{tabular}{ccccccccc}
	$E$ & $1^\text{st}$ & $2^\text{nd}$ & $3^\text{rd}$ & $4^\text{th}$ & $5^\text{th}$ & $6^\text{th}$ & $a_\text{rel}$ & $\mathcal{S}_\text{rel}$\\
	\hline
	\SIrange{5}{10}{MeV}	& 39.6 & 43.4 & 36.8 & 43.7 & 35.8 & 42.1 & 0.07 & 0.28\\
	\SIrange{10}{15}{MeV}	& 57.8 & 64.9 & 54.3 & 65.9 & 52.5 & 63.6 & 0.08 & 0.38\\
	\SIrange{15}{20}{MeV}	& 46.7 & 54.7 & 44.0 & 56.6 & 42.7 & 54.8 & 0.11 & 0.48\\
	\SIrange{20}{25}{MeV}	& 28.2 & 34.7 & 26.4 & 36.8 & 25.9 & 35.8 & 0.14 & 0.48\\
	$>\SI{25}{MeV}$		& 25.1 & 34.2 & 23.0 & 38.3 & 23.1 & 37.4 & 0.21 & 0.71\\
	\hline
	total signal			& 204  & 239  & 191  & 249  & 186  & 241  & 0.11 & 1.00
	\end{tabular}
	\end{center}
	\caption{Event rate per \SI{1}{ms} bin in the Hyper-Kamiokande detector for the six extrema in the time range from \SIrange{200}{235}{ms}, relative amplitude of the SASI oscillations and relative signal-to-noise ratio in the NFS case.
	We exclude the lowest-energy bin ($<$\SI{5}{MeV}) due to the very low event rate.}
	\label{tab-or-extrema}
\end{table}

\begin{figure}[htbp]
	\centering
	\includegraphics[scale=0.49]{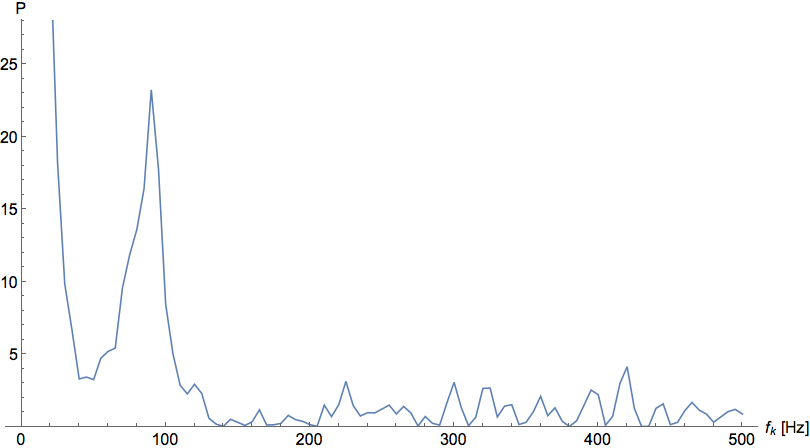}
	\caption{Power spectrum of the total event rate in Hyper-Kamiokande for the NFS case.}
	\label{fig-or-power-spectrum}
\end{figure}
\begin{figure}[htbp]
	\centering
	\includegraphics[scale=0.49]{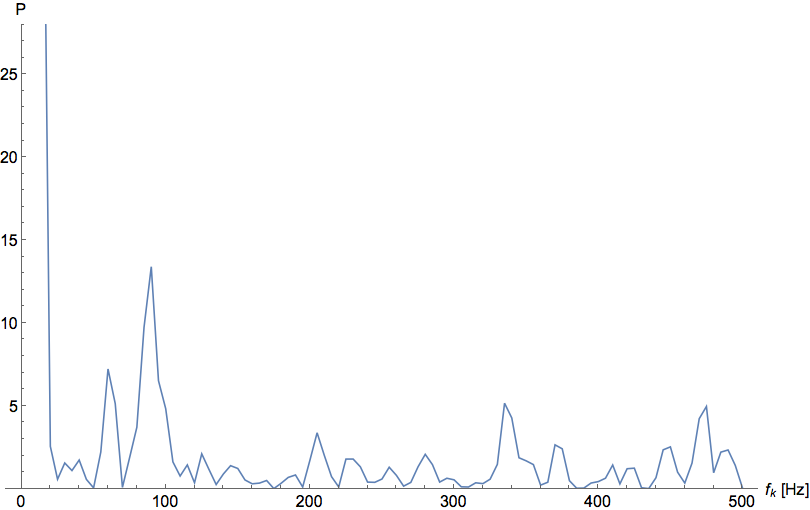}
	\caption{Power spectrum of the $>$\SI{25}{MeV} energy bin in Hyper-Kamiokande for the NFS case.}
	\label{fig-or-power-spectrum-highE}
\end{figure}

We find that the relative amplitude has a strong dependence on the energy bin, with the amplitude in the highest-energy bin ($>$\SI{25}{MeV}) being three times as large as in the low energy bin (\SIrange{5}{10}{MeV}) and nearly twice as large as in the total signal. However, due to the much smaller event count and the corresponding larger relative bin-to-bin fluctuations, the signal-to-noise ratio of the highest-energy bin is still lower than that of the total signal. By adding up multiple energy bins, we can see that $\mathcal{S}_\text{rel}(20, \infty)$ and $\mathcal{S}_\text{rel}(15, \infty)$ are closer to but still less than 1. In appendix~\ref{ap-srel-combine}, we discuss the result of combining two bins.

To cross-check this result, we calculate the power spectrum of the total event rate (figure~\ref{fig-or-power-spectrum}). As expected from figure~\ref{fig-or-snr}, the SASI peak is about equally strong as in the IceCube detector (see figure~\ref{fig-data-power-spectrum}).
When comparing the power spectrum of the total event rate to the power spectrum of the highest-energy bin (figure~\ref{fig-or-power-spectrum-highE}), we see that, while the SASI peak at \SIrange{80}{100}{Hz} is clearly identifiable in both, it is more prominent in the power spectrum of the total event rate.

\subsection{Full Flavor Swap}

\begin{figure}[ht]
	\centering
	\includegraphics[scale=0.5]{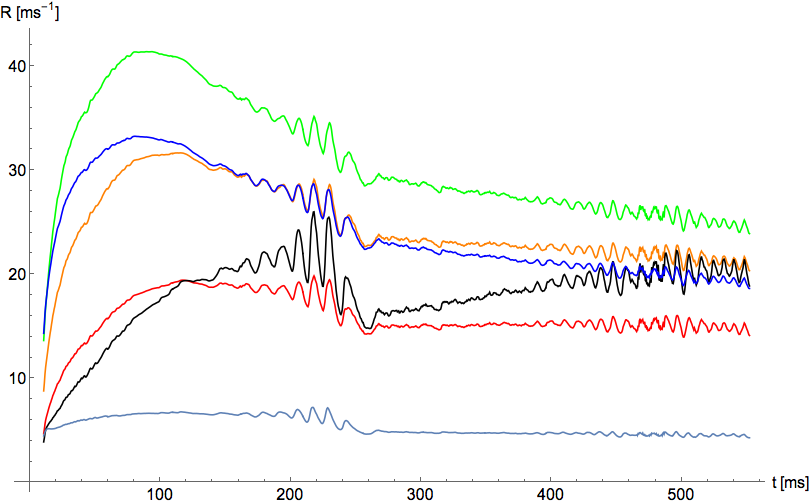}
	\caption{Comparison of event rates for different energy bins in the FFS case: $<$\SI{5}{MeV} (gray-blue), \SIrange{5}{10}{MeV} (blue), \SIrange{10}{15}{MeV} (green), \SIrange{15}{20}{MeV} (orange), \SIrange{20}{25}{MeV} (red), $>$\SI{25}{MeV} (black)}
	\label{fig-or-rate-5mev-bins-x}
\end{figure}

In the other extreme case, full flavor swap, all \nuebar\ detected on Earth originate as \nux\ in the supernova. Compared to the NFS case, in the time interval with the strongest SASI oscillations the neutrino number flux will be about a third lower, while mean energies during the same time will be equal or slightly higher, as shown in figures~\ref{fig-sn-luminosity} and~\ref{fig-sn-mean-energy}.

In figure~\ref{fig-or-rate-5mev-bins-x}, we compare the event rates in \SI{5}{MeV} bins. As in the NFS case, we see that the SASI oscillations in all bins are in phase. Overall, the amplitude is lower than in the NFS case, but there are still notable differences between the relative amplitudes in various energy bins.

In table~\ref{tab-or-extrema-x}, we display the six extrema in the time interval from \SIrange{200}{235}{ms} (the interval where the SASI oscillations are strongest). Additionally, we display the relative amplitude~$a_\text{rel}$ and the relative signal-to-noise ratio~$\mathcal{S}_\text{rel}$.
We find that the relative amplitude in the highest-energy bin is largest, being about twice as large as in the overall signal and about three times as large as in the lowest-energy bins.
However, due to the much smaller event count and the corresponding larger relative bin-to-bin fluctuations, the signal-to-noise ratio of the highest-energy bin is once again lower than that of the overall signal.
When adding up multiple energy bins, we find that $\mathcal{S}_\text{rel}$ increases but does not exceed the value~1. In appendix~\ref{ap-srel-combine}, we discuss the reason for this result.

\begin{figure}[h!]
	\centering
	\includegraphics[scale=0.5]{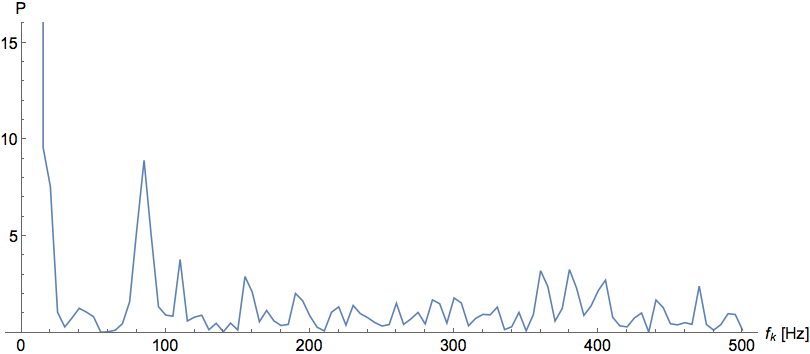}
	\caption{Power spectrum of the total event rate in Hyper-Kamiokande for the FFS case.}
	\label{fig-or-power-spectrum-x}
\end{figure}

\begin{figure}[h!]
	\centering
	\includegraphics[scale=0.5]{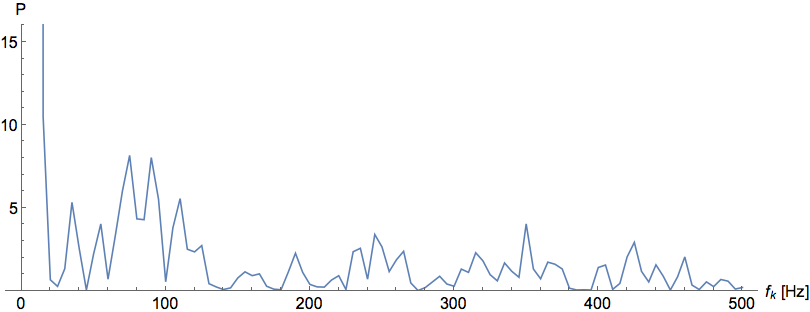}
	\caption{Power spectrum of the $>$\SI{25}{MeV} energy bin in Hyper-Kamiokande for the FFS case.}
	\label{fig-or-power-spectrum-highE-x}
\end{figure}

In figures~\ref{fig-or-power-spectrum-x} and~\ref{fig-or-power-spectrum-highE-x} we show the power spectrum of the total event rate and the highest-energy bin, respectively. While both show a peak at the SASI frequency of \SIrange{80}{100}{Hz}, the peak is not as prominent as in the NFS case, which can be explained by the lower overall event rate and the smaller amplitude of the variations in the event rate.
Because of the relatively large shot noise in the highest-energy bin, the SASI peak may even be spread out to a point where it is hardly possible to determine the precise SASI frequency.
\begin{table}[bhtp]
	\begin{center}
	\begin{tabular}{ccccccccc}
	$E$ & $1^\text{st}$ & $2^\text{nd}$ & $3^\text{rd}$ & $4^\text{th}$ & $5^\text{th}$ & $6^\text{th}$ & $a_\text{rel}$ & $\mathcal{S}_\text{rel}$\\
	\hline
	\SIrange{5}{10}{MeV}	& 27.1 & 28.6 & 25.9 & 28.7 & 25.3 & 28.2 & 0.04 & 0.32\\
	\SIrange{10}{15}{MeV}	& 33.5 & 35.0 & 32.3 & 35.2 & 31.6 & 34.6 & 0.04 & 0.30\\
	\SIrange{15}{20}{MeV}	& 27.0 & 28.7 & 25.4 & 29.1 & 25.4 & 28.6 & 0.05 & 0.39\\
	\SIrange{20}{25}{MeV}	& 17.6 & 19.3 & 16.9 & 19.9 & 16.5 & 19.5 & 0.07 & 0.42\\
	$>\SI{25}{MeV}$		& 20.3 & 24.3 & 19.2 & 26.0 & 18.8 & 25.5 & 0.13 & 0.87\\
	\hline
	total signal			& 132.0  & 143.1  & 127.1  & 146.2  & 124.0  & 143.7  & 0.06 & 1.00
	\end{tabular}
	\end{center}
	\caption{Event rate per \SI{1}{ms} bin in the Hyper-Kamiokande detector for the six extrema in the time range from \SIrange{200}{235}{ms}, relative amplitude of the SASI oscillations and relative signal-to-noise ratio in the FFS case.
	We exclude the lowest-energy bin ($<$\SI{5}{MeV}) due to the very low event rate.}
	\label{tab-or-extrema-x}
\end{table}


\subsection{Larger Oscillation of the Mean Energy}

\begin{figure}[htbp]
	\centering
	\includegraphics[scale=0.5]{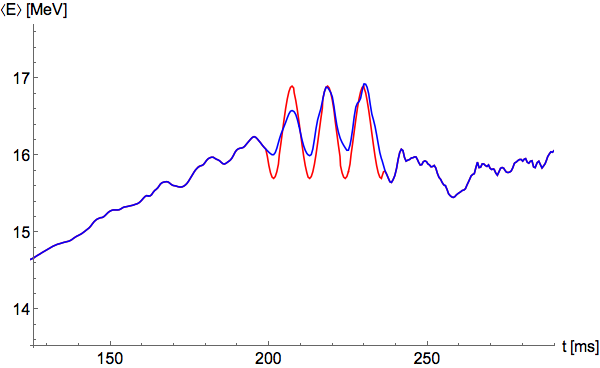}
	\caption{Mean energy (in \si{MeV}) of \nuebar\ in the original data set (blue) and in the LO case (red).}
	\label{fig-or-meane-sin}
\end{figure}
\begin{figure}[htbp]
	\centering
	\includegraphics[scale=0.5]{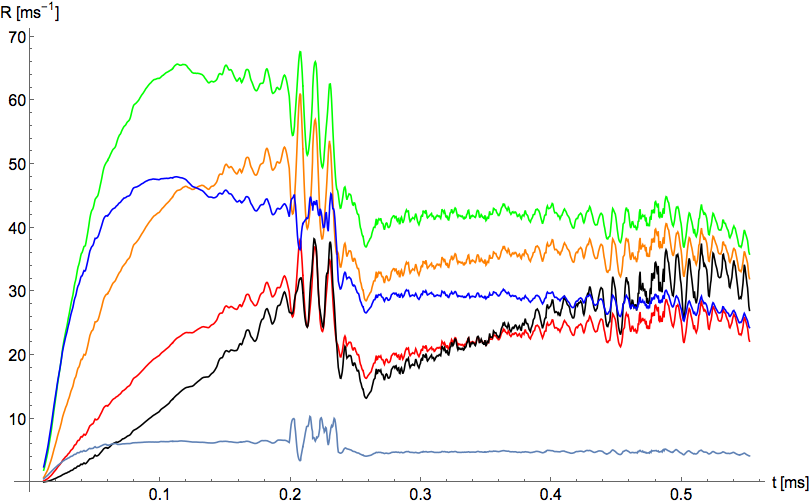}
	\caption{Comparison of event rates for different energy bins in the LO case: $<$\SI{5}{MeV} (gray-blue), \SIrange{5}{10}{MeV} (blue), \SIrange{10}{15}{MeV} (green), \SIrange{15}{20}{MeV} (orange), \SIrange{20}{25}{MeV} (red), $>$\SI{25}{MeV} (black)}
	\label{fig-or-rate-5mev-bins-sin}
\end{figure}
Finally, we want to examine the dependence of these results on the amplitude of the oscillations of the mean energy. To do this, we model the mean energy in the time interval from \SIrange{200}{235}{ms} as a sinusoidal wave that is in phase with the original SASI oscillation but has an increased amplitude. Outside of this time interval, we use the original data, which is displayed in figure~\ref{fig-sn-mean-energy}. The resulting mean energy is shown in figure~\ref{fig-or-meane-sin}. Apart from this modification, we analyze the data as described before. We look at the NFS case only; the results for the FFS case are similar.

As before, we start by splitting the signal into energy bins with a width of \SI{5}{MeV} (see figure~\ref{fig-or-rate-5mev-bins-sin}). This time, we find that the two lowest-energy bins ($<$\SI{5}{MeV} and \SIrange{5}{10}{MeV}) are out of phase. At \SI{206}{ms}, where the other bins have a maximum, these two bins have a minimum; during the following \SI{30}{ms}, no regular oscillations are visible.

In table~\ref{tab-or-extrema-sin}, we list the values of the extrema for the higher-energy bins where the oscillations are in phase. We also show the relative amplitude $a_\text{rel}$ and the relative signal-to-noise ratio $\mathcal{S}_\text{rel}$.

In each of the \SI{5}{MeV} bins, the amplitude is larger than in the total signal. Due to the lower event rate, the detectability in each bin is still worse than in the total signal.
However, combining the three highest-energy bins is obviously advantageous, since these three bins are in phase and have a relative amplitude that is roughly equal. We find that the resulting event rate is nearly half the total event rate and the signal-to-noise ratio improves upon that of the total signal.

Using the condition~\eqref{eq-ap-a2} derived in appendix~\ref{ap-srel-combine}, we can check whether it is helpful to combine the $>$\SI{15}{MeV} bin with the \SIrange{10}{15}{MeV} bin, where the relative amplitude of the SASI oscillations is one third lower. From table~\ref{tab-or-extrema-sin}, we see that, during the time interval of SASI oscillations, the event rate (and thus the event count) in the \SIrange{10}{15}{MeV} bin is about half that of the $>$\SI{15}{MeV} bin. As a result, the condition takes the form
$$
0.12 > 0.18 \cdot 2 \left( \sqrt{1+\frac{1}{2}} - 1\right) \approx 0.08
$$
and is clearly fulfilled.
We calculate $\mathcal{S}_\text{rel} (10, \infty) = 1.23 > \mathcal{S}_\text{rel} (15, \infty)$, in agreement with the prediction.

\begin{table}[bthp]
	\begin{center}
	\begin{tabular}{ccccccccc}
	$E$ & $1^\text{st}$ & $2^\text{nd}$ & $3^\text{rd}$ & $4^\text{th}$ & $5^\text{th}$ & $6^\text{th}$ & $a_\text{rel}$ & $\mathcal{S}_\text{rel}$\\
	\hline
	\SIrange{10}{15}{MeV}	& 54.5 & 67.7 & 51.4 & 66.2 & 49.4 & 62.6 & 0.12 & 0.54\\
	\SIrange{15}{20}{MeV}	& 42.1 & 61.1 & 39.8 & 57.1 & 38.3 & 53.6 & 0.18 & 0.73\\
	\SIrange{20}{25}{MeV}	& 26.0 & 38.2 & 24.5 & 37.0 & 23.7 & 35.1 & 0.20 & 0.64\\
	$>\SI{25}{MeV}$		& 26.5 & 32.2 & 24.4 & 38.3 & 24.4 & 37.8 & 0.18 & 0.59\\
	\hline
	$>\SI{15}{MeV}$		& 94.6 &131.5& 88.7 &132.4& 86.4 &126.5& 0.18 & 1.14\\
	\hline
	total signal			& 204  & 239  & 191  & 249  & 186  & 241  & 0.11 & 1.00
	\end{tabular}
	\end{center}
	\caption{Event rate per \SI{1}{ms} bin in the Hyper-Kamiokande detector for the six extrema in the time range from \SIrange{200}{235}{ms}, relative amplitude of the SASI oscillations and relative signal-to-noise ratio in the LO case. We exclude the low-energy bins, where the oscillations are out of phase.}
	\label{tab-or-extrema-sin}
\end{table}

\begin{figure}[htbp]
	\centering
	\includegraphics[scale=0.5]{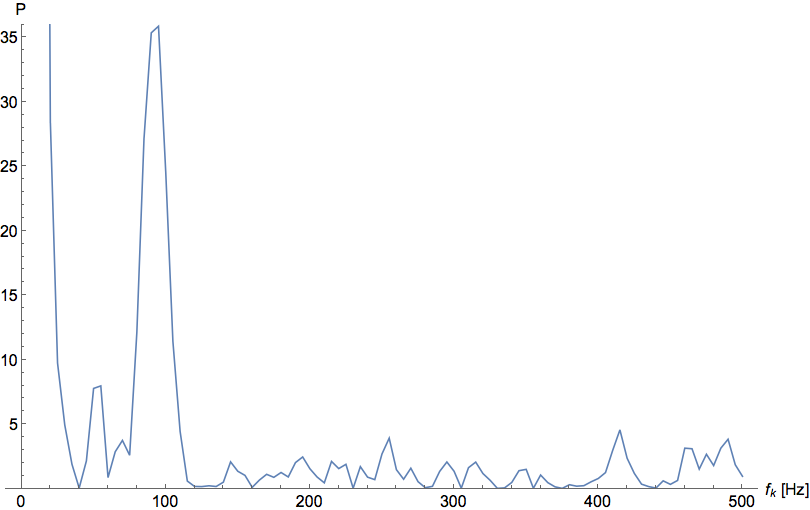}
	\caption{Power spectrum of the $>$\SI{10}{MeV} energy bin in the Hyper-Kamiokande detector in the LO case.}
	\label{fig-or-power-spectrum-highE-sin}
\end{figure}
In figure~\ref{fig-or-power-spectrum-highE-sin} we show the power spectrum of the $>$\SI{10}{MeV} energy bin. We see that the SASI peak is very pronounced, comparing favorably with the peak in the power spectrum of the total signal, which is displayed in figure~\ref{fig-or-power-spectrum}.

\chapter{Conclusions and Outlook}\label{ch-conclusionsoutlook}

While the explosion mechanism of core-collapse supernovae is understood in principle, many details remain unclear. Computer simulations often give conflicting results and are plagued by systematic uncertainties, since out of the three desirable main features of a simulation --~three-dimensionality, detailed treatment of neutrino interactions, full inclusion of General Relativity~-- at most two are currently feasible in any one simulation.
In the foreseeable future, experimental confirmation can only come from a measurement of the neutrino flux from the next galactic supernova. Since galactic supernovae are expected to occur just a few times per century on average, the next one would be a once-in-a-lifetime opportunity.

In this thesis, we tried to find an analysis technique that improves our chances to detect fast time variations in the supernova neutrino flux.
As an example, we used the data set from a three-dimensional simulation by the Garching group, which exhibits clearly visible oscillations in the neutrino number flux and the mean neutrino energy. These oscillations are evidence for the standing accretion shock instability (SASI), which is a hotly debated feature that appears in some simulations but is absent in others.

After describing the current state of knowledge concerning supernova and neu\-tri\-no physics, we showed how the neutrino flux would appear in IceCube. We then explored whether Hyper-Kamiokande can improve upon IceCube’s capabilities by taking advantage of the event-by-event energy information that is available in Hyper-Kamiokande but not in IceCube.

We split the signal into energy bins with a width of \SI{5}{MeV} and found that the SASI oscillations are in phase across all bins. The amplitude of the oscillations varies with energy and is particularly large in the highest-energy bin.
The increased amplitude is counteracted by the lower event rate, which causes an increase in the shot noise. In the simulation that we analyzed here, we found that the latter effect dominates, making it disadvantageous to split the signal into energy bins.

From the power spectrum of the signal, we derived a simple indicator of the signal strength and a condition for when it is advantageous to look at a subset of the signal.
Finally, we showed that a slight increase in the amplitude of the oscillations of the mean energy is sufficient to fulfill this condition. Whether this case is realized in nature remains to be seen.

Looking ahead, the future of supernova research is exciting.
Due to improvements in both software and hardware, the next few years will see sophisticated three-dimensional simulations become more and more commonplace.
On the experimental side, the research community is waiting eagerly for the next galactic supernova. In the meantime, supported by a strong interest in other areas of neu\-tri\-no physics, many new detectors (like Hyper-Kamiokande) and improved detection techniques (like the use of gadolinium to capture free neutrons from inverse beta decay) are being proposed or implemented.
The community is well-prepared for the next galactic supernova and intends to make the most of it and observe it in as many channels as possible~-- neutrinos, the very early light curve with help from SNEWS, and hopefully even gravitational waves.

SN1987A netted only two dozen neutrino events, but confirmed the rough picture that had existed before and led to giant steps forward for neutrino astrophysics.
Thanks to newer, much larger detectors, the observation of a future galactic supernova will net several orders of magnitude more events, bringing a wealth of data to this very data-starved field. Not only would this result in a flurry of activity in the field and answer many remaining questions about supernovae, it could also give new impulses to many related areas of research, be it particle physics at very high energies or stellar nucleosynthesis. If speculations about an unknown energy loss channel are confirmed, it may even open the doors to go beyond today’s standard model of particle physics.

\appendix
\chapter{Calculation of the Detection Rate}

\section{Energy-Dependent IBD Cross Section}\label{ap-IBD-cross-section}
For the energy dependence of the inverse beta decay (IBD) cross section, $\tdiffx{\sigma}{E_e}$, we used a formula derived by Strumia and Vissani~\cite{Strumia2003}. In this section, we will extract the equations that are necessary for the calculations.
These equations will use the definitions
\begin{align}
s - m_p^2	&= 2 m_p E_\nu\\
s - u		&= 2 m_p (E_\nu + E_e) - m_e^2\\
t		&= m_n^2 - m_p^2 - 2 m_p (E_\nu - E_e)\\
f_1		&= \frac{1 - (1 + \xi) t / (4M^2)}{[1 - t / (4M^2)] (1 - t / M_V^2)^2}\\
f_2		&= \frac{\xi}{[1 - t / (4M^2)] (1 - t / M_V^2)^2}\\
g_1		&= \frac{g_1(0)}{(1 - t / M_A^2)^2}\\
g_2		&= \frac{2 M^2 g_1}{m_\pi^2 - t},
\end{align}
with the constants $\Delta = m_n - m_p \approx \SI{1.293}{MeV}$, $M = (m_p + m_n) / 2 \approx \SI{938.9}{MeV}$, $M_A^2 \approx \SI{1}{GeV^2}$, $M_V^2 \approx \SI{0.71}{GeV^2}$, $g_1 (0) = -1.27$ and $\xi = 3.706$.

In its most compact form, the energy dependent IBD cross section can be written~as 
\begin{equation}
\tdiff{\sigma (E_\nu, E_e)}{E_e} = 2 m_p \tdiff{\sigma}{t}\quad \text{if}\quad E_\nu \geq E_\text{thr} = \frac{(m_n + m_e)^2 - m_p^2}{2 m_p},
\end{equation}
where $E_\text{thr}$ is the threshold energy for IBD.
$\tdiffx{\sigma}{t}$ is given by
\begin{equation}
\tdiff{\sigma}{t} = \frac{G_F^2 \cos^2{\theta_C}}{2 \pi (s - m_p^2)^2} \abs{\mathcal{M}^2}.
\end{equation}

The matrix element $\abs{\mathcal{M}^2}$ can be calculated to be
\begin{align}
\abs{\mathcal{M}^2} &= A(t) - (s-u) B(t) + (s-u)^2 C(t),
\end{align}
where
\begin{align}
A(t) &\approx M^2 (f_1^2 - g_1^2)(t - m_e^2) - M^2 \Delta^2 (f_1^2 + g_1^2) - 2 m_e^2 M \Delta g_1 (f_1 + f_2)\\
B(t) &\approx t g_1 (f_1 + f_2)\\
C(t) &\approx \frac{f_1^2 + g_1^2}{4}
\end{align}
are the NLO approximations instead of their more complicated precise values. This introduces an error of typically below 1\,\%, depending on the neutrino energy.

\section{Energy-Dependent \nuebar\ Flux}\label{ap-nu-flux}
The energy-dependent \nuebar\ flux $\tdiffx{\Phi_\nu}{E_\nu}$ in the detector can be written as
\begin{equation}
\tdiff{\Phi_\nu (E_\nu, d, t)}{E_\nu} = \frac{1}{4 \pi d^2} \frac{L(t)}{\langle E_\nu \rangle (t)} f (E_\nu, t),
\end{equation}
where $d$ is the distance between the supernova and the detector, $L$ is the neutrino luminosity of the supernova in the observer direction, $\mean{E_\nu}$ is the mean neutrino energy, and $f$ is the spectral distribution of neutrino energies. It can be approximated by a normalized Gamma distribution~\cite{Keil2003,Tamborra2012}:
\begin{equation}
f (E_\nu) = \frac{E_\nu^\alpha}{\Gamma (\alpha + 1)} \left( \frac{\alpha + 1}{A} \right)^{\alpha + 1} \exp \left[ - \frac{(\alpha + 1) E_\nu}{A} \right]
\end{equation}
In this formula, $A$ is an energy scale, while $\alpha$ determines the shape of the distribution: $\alpha = 2$ corresponds to a Maxwell-Boltzmann distribution, while $\alpha > 2$ corresponds to a “pinched” spectrum, which is more typical for neutrino spectra from supernovae.

The first two energy moments of the distribution are
\begin{align}
\mean{E_\nu}		&= \int_0^\infty \d E_\nu\, E_\nu f(E_\nu) = A\\
\mean{E_\nu^2}	&= \int_0^\infty \d E_\nu\, E_\nu^2 f(E_\nu) = \frac{\alpha + 2}{\alpha + 1} A^2,
\end{align}
and therefore,
\begin{equation}
\alpha = \frac{\mean{E_\nu^2} - 2 A^2}{A^2 - \mean{E_\nu^2}}.
\end{equation}
\enlargethispage{\baselineskip} 

Both the mean energy and the mean squared energy are part of the data set described in chapter~\ref{ch-data-dataset}. This enables us to calculate the Gamma distribution, and thus the flux, from these formulas.

\chapter{Calculation of the Shot Noise Power}\label{ap-shot-noise}
In order to estimate the shot noise in the Hyper-Kamiokande detector, we consider a signal consisting of a sequence $t_{1,…,N}$ of measured arrival times, which sample the rate $R(t)$ over the signal duration $\tau$. The Fourier transform of this signal is
\begin{equation}
g(f) = \int_0^\tau \d t\, R(t) e^{-i 2 \pi f t} = \sum_{j=1}^{N} e^{-i 2 \pi f t_j} \label{eq-ap-ft}
\end{equation}
and the spectral power is $G(f) = \abs{g(f)}^2$.

\begin{figure}[htbp]
	\centering
	\includegraphics[scale=0.5]{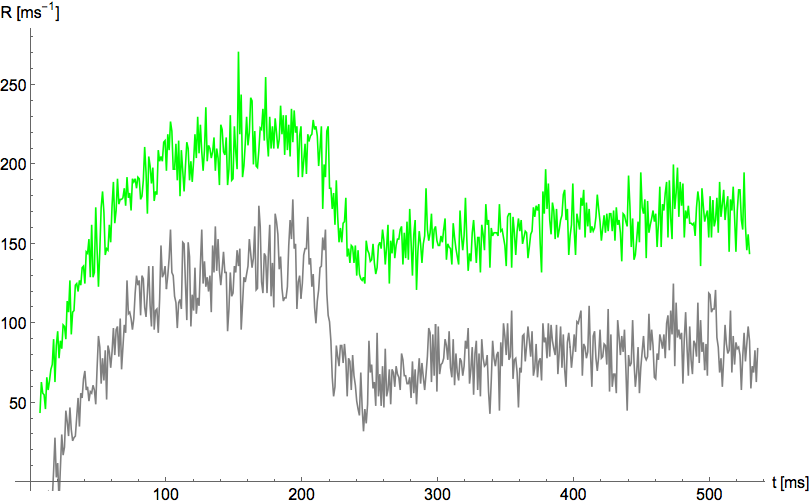}
	\caption{The “equivalent flat” signal (green). For comparison, we also show the regular \SI{1}{ms} bin event rate shifted down by \SI{80}{ms^{-1}} (gray).}
	\label{fig-ap-ef-signal}
\end{figure}

If we choose the “equivalent flat” neutrino signal introduced in chapter~\ref{ch-data-normalization} (see figure~\ref{fig-ap-ef-signal}), the value of $2 \pi f t_j$ (modulo $2 \pi$) is distributed approximately uniformly. 
The sum in equation \eqref{eq-ap-ft} can therefore be considered a random walk with unit step size in the complex plane~\cite{Lund2010,Dighe2003}.
Accordingly, the ensemble average is independent from frequency for $f \neq 0$ and follows the normalized exponential distribution $p(G) = N^{-1} e^{-G/N}$, which has the expectation value $\mean{G} = N$.

For $f = 0$ meanwhile, the Fourier transform is $g(0) = \sum_{j=1}^N 1 = N$ and the spectral power is $G(0) = N^2$. Accordingly,
\begin{equation}
\frac{ \mean{G_{f \neq 0}} }{ G_{f=0} } = \frac{1}{N} .
\end{equation}
Introducing the Hann window function defined in equation \eqref{eq-data-hann} modifies the expectation value of the spectral power to be $\mean{G} = \mean{w^2} N$, where $\mean{w^2} = 3/2$~\cite{Lund2010}.

Switching now to the variables defined in chapter~\ref{ch-data-ft}, that were binned and multiplied by the Hann function, we define the shot noise as the frequency-averaged spectral power, excluding $f=0$ and $f = f_\text{max}$.

We find that
\begin{align*}
P_\text{shot} = \mean{P(f_k)} &= \frac{2 \mean{\abs{h(f_k)}^2} }{N_\text{bins}^2}\\
&= \frac{2}{N_\text{bins}^2} \cdot \frac{ \mean{\abs{h(f_k)}^2} }{\abs{h(f_0)}^2} \cdot \abs{h(f_0)}^2\\
&= \frac{2}{N_\text{bins}^2} \cdot \frac{3}{2} \frac{1}{N_\text{events}} \cdot N_\text{events}^2\\
&= \frac{3 N_\text{events}}{N_\text{bins}^2}.
\end{align*}

It is important to note that in this approximation, the shot noise only depends on the total number of events and the number of bins, both of which remained unchanged by the multi-bin averaging that brought upon the “equivalent flat” signal.

\chapter{The Relative Signal-to-Noise Ratio $\mathcal{S}_\text{rel}$}

\section{Derivation of $\mathcal{S}_\text{rel}$ From the Power Spectrum}\label{ap-srel-derive}

Let us assume that the total signal consists of $N_\text{tot}$ events and contains SASI oscillations with a relative amplitude $a_\text{tot}$. The minimum relative amplitude needed for the oscillations to be detectable in the power spectrum is $a_\text{min, tot} = \sqrt{6/N_\text{tot}}$, according to equation~\eqref{eq-data-amin}.

Furthermore, let us assume that a subset of that signal consists of $N_\text{sub}$ events and contains SASI oscillations with a relative amplitude $a_\text{sub}$. The minimum relative amplitude needed for the oscillations to be detectable in the power spectrum is $a_\text{min, sub} = \sqrt{6/N_\text{sub}}$.

The detectability of the SASI oscillation in that subset is better than in the total signal, if
\begin{equation}
\frac{a_\text{sub}}{a_\text{min, sub}} > \frac{a_\text{tot}}{a_\text{min, tot}}.
\end{equation}
Inserting the formulas for $a_\text{min}$, we have
\begin{align}
a_\text{sub} \sqrt{N_\text{sub}} &> a_\text{tot} \sqrt{N_\text{tot}}\\
\frac{a_\text{sub}}{a_\text{tot}} \sqrt{\frac{N_\text{sub}}{N_\text{tot}}} &> 1. \label{eq-ap-a}
\end{align}

If the subset is an energy bin,
\begin{align}
N_\text{sub} &= \int \d t\, R_\text{sub} (t)\\
	&= \iint \d t\, \d E\, R_\text{tot} (t) f(E,t)\\
	&\approx \iint \d t\, \d E\, R_\text{tot} (t) f(E)\\
	&= N_\text{tot} \int \d E\, f(E, t)
\end{align}
where $R_\text{sub}$ and $R_\text{tot}$ are the event rates of the subset and the total signal, respectively, and $f(E)$ is the time-average of $f(E,t)$. Since $f(E,t)$ typically varies by less than 10\,\% during the time interval where the SASI oscillations are strongest, this is a good approximation.

Inserting this approximation of $N_\text{sub}$ into the condition~\eqref{eq-ap-a}, we get
\begin{equation}
\frac{a_\text{sub}}{a_\text{tot}} \sqrt{ \int \d E\, f(E, t)} > 1.
\end{equation}

If we choose the subset to be all events with an energy between $E_1$ and $E_2$, this condition is equivalent to $\mathcal{S}_\text{rel} (E_1, E_2) > 1$.

\section{The Combined $\mathcal{S}_\text{rel}$ of Multiple Bins}\label{ap-srel-combine}
Let us assume that we have two energy bins, which are disjoint subsets of a total signal. Each bin $i$ consists of $N_i$ events, contains SASI oscillations with a relative amplitude $a_i$ and has a relative signal-to-noise ratio $\mathcal{S}_i$. We also assume that the SASI oscillations are in phase across the two bins. Furthermore, without loss of generality we let $\mathcal{S}_1 \geq \mathcal{S}_2$.

The combination of both bins then consists of $N_{1+2} = N_1 + N_2$ events, contains SASI oscillations with a relative amplitude
\begin{equation}
a_{1+2} = \frac{a_1 N_1 + a_2 N_2}{N_1 + N_2}
\end{equation}
and, as discussed in appendix~\ref{ap-srel-derive}, its relative signal-to-noise ratio can be written as
\begin{equation}
\mathcal{S}_{1+2} = \frac{a_{1+2}}{a_\text{tot}} \sqrt{\frac{N_{1+2}}{N_\text{tot}}}.
\end{equation}

Combining both bins increases the detectability if $\mathcal{S}_{1+2} > \mathcal{S}_1$. Solving this condition for $a_2$, we find:
\begin{align}
\frac{a_1 N_1 + a_2 N_2}{N_1 + N_2} \sqrt{N_1 + N_2} &> a_1 \sqrt{N_1}\\
a_1 N_1 + a_2 N_2 &> a_1 \sqrt{N_1 (N_1 + N_2)}\\
a_2 N_2 &> a_1 N_1 \left( \sqrt{1+ \frac{N_2}{N_1}} - 1 \right)\\
a_2 &> a_1 \frac{N_1}{N_2} \left( \sqrt{1+ \frac{N_2}{N_1}} - 1 \right) \label{eq-ap-a2}
\end{align}

In the limiting case where $N_2 \ll N_1$,
$$
\sqrt{1+ \frac{N_2}{N_1}} \approx 1 + \frac{N_2}{2 N_1}
$$
and the condition~\eqref{eq-ap-a2} simplifies to
\begin{equation}
a_2 > \frac{a_1}{2}.
\end{equation}

Put another way, if the relative amplitude in a bin with a very low event rate is at least half of the relative amplitude of the total signal, substracting that bin from the total signal will not increase $\mathcal{S}_\text{rel}$. Accordingly, in the NFS and FFS cases, where the relative amplitude of the SASI oscillations in the lowest-energy bin is greater than half the relative amplitude of the total signal, disregarding these bins will not improve the detectability of SASI oscillations.

\backmatter
\bibliography{thesis}
\chapter*{Acknowledgements}
\thispagestyle{empty}

First and foremost, my sincere thanks go to Georg Raffelt for welcoming me in his group and suggesting this topic. The many discussions and his overall support have been of inestimable value.

I would like to thank Irene Tamborra for providing me with the preprocessed data set and for several valuable discussions, and Lothar Oberauer, who has agreed to act as my supervisor.

I am grateful to my colleagues at the Max-Planck-Institut and to the organizers, lecturers and participants of the 2014 NBIA PhD-School {\em Neutrinos underground \& in the heavens} for many interesting discussions and fond memories.

My friends and family have given me so much love, support and chocolate while I worked on this thesis. Thank you.

Finally, I am indebted to the many giants upon whose shoulders I now stand.

\end{document}